# Comparing wins above replacement across baseball history

Michael Doane
Independent Researcher, Cary, North Carolina, USA
ORCID: 0009-0003-0521-8981

Abstract

This paper introduces iWAR, a system for estimating wins above replacement from historical Retrosheet play-by-play records. We compare its runs-allowed (RA) and fielding-independent pitching (FIP) versions with Baseball-Reference WAR (bWAR), FanGraphs WAR (fWAR) and run-based alternatives. The common historical sample contains 1,471 team-seasons across 79 years and 1,441 consecutive franchise pairs. Under the fixed conversion to wins, same-season mean absolute error was 3.695 credited wins per 162 games for iWAR RA, compared with 4.724 for bWAR and 4.744 for fWAR: reductions of 21.793% and 22.112%. The paired differences were −1.030 and −1.049, with conditional 95% calendar-block intervals of [−1.367, −0.706] and [−1.398, −0.725]. Calibration preserved most of these advantages. A broader bWAR comparison across 118 years retained a smaller advantage; modern differences were smaller or sometimes reversed. PythagenPat reconstructed records more closely than every WAR total. Next-season replay gave nearly identical fixed-map errors for iWAR FIP and fWAR, while calibration changed their ordering. An amended chronological learned-model analysis of 1,552 franchise pairs gave neural RMSE 10.163, versus 9.946 for the development-selected ridge comparator (difference +0.217; conditional interval [0.015, 0.474]). This work provides a new, specialized framework for analyzing WAR across eras and provides context for future work using learning models.

## 1 Introduction

A baseball team wins 95 games, and its players accumulate a combined WAR total. How closely does a team's total WAR, added to the wins expected from a replacement-level team over the same number of games, reproduce its actual win total? Would the same measure also tell us how well the franchise will perform next year? These are natural questions for a statistic expressed in wins. They are also different questions, and the distinction becomes consequential when competing implementations are compared across a century of baseball.

Wins above replacement estimates a player's contribution relative to a replacement-level alternative (Slowinski, 2010). Its appeal is straightforward: hitting, pitching and fielding can be discussed on a common scale. The calculation is less straightforward. Each implementation must decide how recorded events become run values, how to adjust for context, how to distribute responsibility among players and how to convert runs into wins. A comparison between WAR measures therefore evaluates a collection of choices. Two measures can agree about the broad ranking of players while differing in the total credit they assign to a team or in the aspects of performance they preserve.

Existing research has made these choices more transparent. Baumer, Jensen and Matthews (2015) developed openWAR using public event data, a conservation-of-runs framework and resampling-based uncertainty estimates. Their work established a direct precedent for an open implementation of overall player value. Brill and Wyner (2024) developed Grid WAR for starting pitchers, showing how nonlinear game-level valuation changes the credit assigned to highly variable performances. Their historical application and predictive comparisons demonstrate that a new WAR measure should be assessed against a clearly specified target. Broader efforts to compare players across eras, including Full House Modeling, address differences in talent pools that a within-season league reference does not estimate (Yan et al., 2025).

Fort, McMichael and Rucker (2025) directly examine team-level bWAR and fWAR in wins-production estimation. Their work establishes prior study of the WAR-to-team-wins question; the present contribution concerns the transparent historical valuation and the matched targets, conversions and coverage examined here. Neural baseball team models also precede this study: Hall and Seaman (2009) analyze team performance, Farrell (2019) presents a student project predicting team wins, and Lee et al. (2020) compare LSTM, ARIMA and Grey winning-percentage models. These studies use different information and evaluations. We therefore compare retained predictions on explicit shared rows rather than import their reported accuracies. Strong tabular comparators are also necessary when evaluating neural models (Grinsztajn, Oyallon and Varoquaux, 2022).

The available detail also changes across baseball history. Total Zone and Fielder Range use historical records to estimate defensive contributions under the information available in their respective sources (Smith, n.d.; Johnson, 2011). Spatial models such as Bayesball use richer observations of fielding opportunities and explicitly pool information across players (Jensen, Shirley and Wyner, 2009). These approaches serve related purposes, but their inputs and reference conditions differ. Historical coverage is therefore part of a comparison's meaning. A result based mainly on early seasons may tell a different story from a result dominated by contemporary baseball.

A second issue concerns the meaning of successful prediction. A measure that accounts closely for a completed season can preserve information that is weakly persistent into the following year. Conversely, a measure designed to suppress transient variation may forecast reasonably while explaining less of the variation that actually occurred. Shmueli (2010) describes the broader distinction between statistical description, explanation and prediction. In baseball, the distinction is especially visible because observed scoring, team wins and future performance are all plausible evaluation targets. A useful comparison must say which of these is being assessed.

We examine these questions using iWAR, a system for estimating wins above replacement from historical Retrosheet play-by-play records, and the observed historical team totals published by Baseball-Reference and FanGraphs. The iWAR calculation has a conventional runs-allowed pitching version, iWAR RA, and a separately specified fielding-independent pitching version, iWAR FIP. The primary comparison puts all four methods on identical team-seasons. It then asks whether their differences persist after each measure receives its own estimated conversion to wins. Finally, a broader comparison with bWAR examines how the findings change when historical seasons missing from the common provider sample become available.

The resulting contribution is a matched historical benchmark of WAR under explicitly different evaluation tasks. Its scope extends from 1908 through 2025, with a common four-method sample of 79 observed years and a secondary bWAR sample spanning all 118 years. The design allows several explanations for an apparent advantage to be examined directly. A poor conversion to wins might create avoidable error. Missing historical periods might alter a pooled result. A measure's same-season success might carry weakly into the next season. Reporting these analyses together makes the consequences visible.

The principal finding is clear. On the common historical sample, iWAR RA reconstructs the completed season more accurately than bWAR or fWAR, reducing mean absolute error by approximately 22%. Most of that difference remains after scalar calibration. The advantage becomes smaller when the bWAR comparison expands, and next-season replay produces a different ordering. These results identify the task for which the observed superiority holds. They also provide a concrete reason to report historical coverage and conversion choices whenever team-level performance is used to compare WAR measures.

## 2 Data and methods

### 2.1 Sources and study populations

Retrosheet event files and game logs supply the underlying plays, team records and scoring totals (Retrosheet, 2026). The source inventory contains 2,594 potential American League and National League team-seasons from 1908 through 2025. Complete conventional iWAR RA and FIP products are available for 2,568; the remaining 26 have incomplete products and are excluded from the complete-season comparison. An additional 14 complete-product team-seasons fail the reconciled-record requirement. Their event and game-log sources agree on the number and identities of games but differ in credited wins. Those discrepancies are retained without an inferred repair. Sections S1 and S3 of the supplement describe the eligibility process and each exclusion.

The external measures are actual values observed through the ordinary public interfaces of Baseball-Reference and FanGraphs. Baseball-Reference's supplied team Total enters once. FanGraphs team batting WAR includes all batting roles, including pitcher batting, and its pitching-only WAR is added once. Independently ranked provider columns are joined by team identity before arithmetic is checked. This procedure avoids treating overlapping batting and pitching categories as separate contributions. Player names, abbreviations and franchise relocations are resolved explicitly; code equality alone does not establish continuity.

All 118 bWAR source years were obtained or preserved. Complete fWAR batting and pitching roles were available in 79 years. The primary comparison requires complete iWAR products, finite supplied provider totals, compatible regular-season scope and reconciled team records. It contains 1,471 team-seasons. The observed years are 1908–1967, 1975, 1988–1997, 2017–2019 and 2021–2025. These years form the achieved sample; the gaps remain part of the reported coverage. Missing provider values are retained as missing and receive no inferred replacement.

The replay analysis pairs an eligible origin-year valuation with the next season's complete franchise record. A pair must connect consecutive calendar years and a verified continuing franchise. Target-year WAR is unnecessary because the predictor comes entirely from the origin year. These rules yield 1,441 pairs in 78 target years. In reporting coverage, origin years identify the source of the predictor, while target years identify the outcomes being scored. This distinction matters around missing-year boundaries.

A secondary comparison removes the requirement for fWAR while retaining the complete iWAR and supplied-bWAR requirements. It contains 2,554 team-seasons across 118 years and 2,524 replay pairs across 117 target years. The secondary protocol was specified after the primary findings were known and frozen before its new losses were computed. We report its original common rows separately from the added history. The broader comparison therefore examines the consequences of available coverage while preserving the primary four-method analysis.

Provider records and their revisions introduce additional limits. Complete iWAR game identities are certified within the retained source system, but provider tables do not publish full game-identity lists. Agreement in reported records and role totals supports compatibility without proving exact provider membership for every game. Historical provider WAR can also change. For example, FanGraphs' 2024 update applied Statcast-based fielding and baserunning changes retroactively to 2016 (Appelman, 2024). Our comparisons use the values observed when the tables were collected. Archived season-end vintages were unavailable.

### 2.2 The conventional iWAR calculation

The iWAR calculation expresses batting, running, fielding, pitching and position contributions in runs, converts them within league-season, and adds replacement credit. Actual event roles keep ordinary batting, pitcher batting and pitching separate. Cross-league stints receive the conversion appropriate to their own league before they are combined in wins. Unknown core inputs remain missing. A component that the available events cannot measure retains its declared assumed-average incremental convention; its numerical treatment should not be read as a measured assessment of the corresponding skill.

Batting weights are derived from recorded same-season events and run expectancy. A base-out state specifies the occupied bases and the number of outs. For a completed event, its observed run value combines runs scored during the event with the change in expected runs remaining in the half-inning. The calculation distinguishes the event category from advancement beyond a stated normal transition. Category values are then averaged over a common league-season distribution of pre-plate-appearance states. Thus the same category receives a common batting weight within a league-season, while departures attributable to certified running opportunities enter the running component. The full state references, shrinkage rules and exclusions are specified in Section S2 of the supplement.

The conventional defensive allocation uses eligible balls in play and active fielding positions. Known initiating fielders and responsible errors contribute to the primary conversion assignment; unresolved responsibility is allocated fractionally and flagged. For each player-position-hand stratum, observed primary conversions are compared with a league reference at the same exposure. The difference is multiplied by a standardized run gap between conversion and nonconversion. This coarse method uses the recorded opportunities available throughout the historical source period. It provides limited information about physical range, arm strength or other specific defensive abilities.

The RA pitching calculation starts from expected runs over the pitcher's actual recorded outs and subtracts charged runs. Expected runs depend on the league run environment and a same-season park factor. Fielding support is then subtracted from pitching credit in the same amount that it is assigned to fielders. A pitcher's own fielding enters once. Inherited-run charges follow the officially responsible pitcher, while defensive support follows the pitcher active during the event. When automatic-runner placements resolve completely, those runners' own scored runs are removed consistently from the charged and league-reference runs. Otherwise, the season retains all runs under an explicitly flagged convention.

The park factor compares adjusted runs per out at a site with road games of its home clubs, shrinks the log ratio according to home-game exposure, and normalizes the pitching-out-weighted league mean to one. It applies only to RA expectation. Batting, fielding and FIP retain neutral component park factors. This is a descriptive same-season adjustment; the estimated factor can reflect the teams, opponents and schedules represented at that park.

The FIP alternative changes the pitching component. Its numerator is 13 times home runs, plus 3 times walks and hit batters, minus 2 times strikeouts. The implementation includes intentional walks and inside-the-park home runs, and uses a league constant anchored to the observed run environment. It does not add infield flies or apply the RA defensive-support subtraction, park factor, or separate leverage adjustment. These choices distinguish it from FanGraphs pitching WAR, which uses park-adjusted FIP with infield flies counted as strikeouts and its own replacement, leverage and conversion conventions (Slowinski, 2012). Baseball-Reference pitching WAR instead begins with runs allowed and adjusts for opposition, parks and team defense (Sports Reference, n.d.-b). The labels RA and FIP identify the pitching basis; they do not imply identical implementations across providers.

Position credit follows a fixed schedule and actual role exposure. The schedule, historical assumptions and fallback rules are retained in the supplement. League-season centering removes the total of each declared role pool in proportion to its corresponding exposure: ordinary plate appearances, pitcher-role plate appearances or pitching outs. These adjustments use the recorded run components and exposures. The scored team's actual win total does not determine its conventional WAR or its centering correction.

Runs above average are converted using a common league-season runs-per-win value,

$$\text{RPW} = 10\sqrt{\frac{R^{\text{adj}}/G_{\text{team}}}{4.5}}.$$

Here $R^{\text{adj}}$ is adjusted league runs and $G_{\text{team}}$ counts each covered game once for each participating team. Replacement contributes 0.206 wins per covered team-game, allocated 57% through ordinary batting exposure and 43% through pitching outs. Consequently, a complete centered league-season satisfies

$$\sum_{\text{teams}} \text{WAR} = 0.206 \sum_{\text{teams}} G.$$

Adding the fixed replacement-win fraction of 0.294 makes implied league wins sum to half of both-teams games. This identity helps explain small aggregate signed bias. It leaves the errors of individual teams free to vary, so it does not determine mean absolute or squared error.

## 2.3 Reconstruction and next-season replay

Credited actual winning fraction is

$$y_i = \frac{W_i + 0.5T_i}{G_i},$$

where a tie contributes half a win and $G_i = W_i + L_i + T_i$. The fixed WAR map is

$$\hat{y}_i = 0.294 + \frac{\text{WAR}_i}{G_i}.$$

For reconstruction, the predictor and outcome refer to the same completed season. For replay, origin-year WAR per origin-year game predicts the next season's credited winning fraction. Future games played never enter the predictor. The primary error is $162(\hat{y}_i - y_i)$, expressed in credited wins per 162 played games. This common unit permits comparisons across seasons with different lengths. It scales the error in a winning fraction; it does not fill gaps in incomplete WAR or annualize missing player contributions.

Run-based references provide a demanding comparison for team-level reconstruction. The exponent-two Pythagorean expectation is $RS^2/(RS^2 + RA^2)$. PythagenPat uses the same form with exponent $\gamma = [(RS + RA)/G]^{0.285}$, fixed from the published convention (Sports Reference, n.d.-a). The general relationship between run distributions and Pythagorean expectations has a mathematical foundation described by Miller (2007). Our inputs are reconciled, unadjusted game-log runs scored and allowed. They remain distinct from the automatic-runner-adjusted references used in parts of iWAR.

A constant .500 prediction is included for both tasks, and previous credited record is included for replay. Each displayed comparison uses identical eligible rows. Where a particular baseline requires a narrower run-input intersection, all WAR methods are rescored on that intersection. Actual-season win errors are supplied alongside the primary per-162 results so that the effect of normalization remains visible.

## 2.4 Estimated conversion to wins

The fixed map has an appealing interpretation, but a measure's scale or compression could affect its error independently of the information it contains. We therefore estimate an intercept and slope separately for each nonconstant predictor:

$$y_i = \alpha + \beta x_i + \varepsilon_i .$$

The WAR predictor is WAR per game. Other predictors are the Pythagorean or PythagenPat winning fraction, run differential per game, and previous record for replay. Ordinary least squares uses equal training-row weights and leaves predictions unclipped. The .500 reference remains fixed. Every method within a comparison receives the same training and evaluation rows.

Same-season calibration leaves out the complete scored year and fits the conversion using the other eligible years. Later years may therefore train an earlier year's conversion. This analysis asks how well the recorded measure maps to completed-season results after correcting its overall level and scale. Its design separates the scored year from estimation of that scalar relationship while retaining the retrospectively developed underlying WAR methods.

Replay calibration uses an expanding window of strictly earlier completed target years. A scored year requires at least ten earlier target years and 150 training pairs. The use of multiple chronological test periods follows the general evaluation principle described by Tashman (2000). The resulting common evaluation subset contains 1,282 pairs in 67 target years. Fixed predictions are rescored on those exact rows, so a change attributed to calibration cannot arise simply from using a different set of teams.

The broader bWAR analysis applies the same rules to its own training population. All 1,450 secondary training cohorts differed from the original ones and received newly estimated scalar coefficients during the completed study. The original 1,089 maps were preserved. An explicit pre-loss amendment specified when a coefficient could be reused; none met that condition in the expanded training populations. No player valuation was refitted for these comparisons.

### 2.5 Errors, historical periods and uncertainty

Mean absolute error (MAE) gives the average size of the win error. Root mean squared error (RMSE) gives greater weight to large misses. Bias is the mean signed error, with positive values indicating overprediction. These summaries answer related questions and can rank methods differently. Their interpretation follows the general requirement to match forecast evaluation with the quantity being assessed (Gneiting, 2011). We report both MAE and RMSE instead of choosing between them after observing the ordering.

Primary pooled results give each team-season or replay pair equal weight. Equal-season weighting is a sensitivity analysis because later leagues contain more teams. Further results separate leagues, restrict season exposure to at least 140 games, and retain every individual year. The fixed reporting periods are 1908–1945, 1946–1968, 1969–1993, 1994–2014 and 2015–2025.

Replay periods use target years. Correlations and prediction $R^2 = 1 - SSE / SST$ appear in the supplement; squared correlation is reported separately and carries a different interpretation.

Paired interval calculations preserve the same methods and teams within each resampled historical block. The primary procedure uses 5,000 draws of nonoverlapping five-year calendar blocks anchored at 1908. Every draw samples the original number of nonempty blocks with replacement and recomputes the paired MAE or RMSE difference. Linear 2.5th and 97.5th percentiles form the interval. At least ten nonempty blocks are required; one-year and ten-year block sensitivities are also retained. The original comparison and secondary expansion use their separately recorded seeds and saved draws.

Block resampling has established precedent for dependent observations (Künsch, 1989). Dependence is particularly consequential when sports observations share teams or outcomes, as illustrated in a different setting by Brill, Yurko and Wyner (2026). Our intervals condition on the fitted measures, observed inputs and saved calibration coefficients. They omit method-development, source-revision and coefficient-fitting uncertainty. They also depend on how well the chosen calendar blocks represent the historical dependence. The term “conditional interval” refers to these fixed-input resampling summaries throughout the paper.

Provider display precision is assessed separately. Team values rounded to one decimal imply bounded perturbations of their fixed-map predictions. Bounds are propagated to aggregate losses and paired effects, using actual origin-game denominators. For calibrated predictions they condition on the estimated coefficients and omit the effect of rounding training inputs. A tiny paired difference should therefore be interpreted alongside both its interval and its display-precision bound.

### 2.6 Accounting and player agreement

To describe the relationship between a WAR prediction and PythagenPat, let $e_R$ be the run-based prediction error and $d$ be the WAR prediction minus the run-based prediction, both in wins per 162. Then $e_{\mathrm{WAR}} = e_R + d$, giving

$$\mathrm{MSE}_{WAR} - \mathrm{MSE}_R = \mathrm{mean}_w\left(d^2\right) + 2\,\mathrm{mean}_w\left(e_R d\right).$$

The first term measures squared distance between the predictions. The second shows whether that distance tends to offset or reinforce the run-based errors. This exact identity describes differences in squared error. A causal allocation to particular WAR components would require additional information, including a complete aligned provider component ledger.

A separate bWAR agreement analysis uses exact player identities from the Chadwick Persons Register (Chadwick Baseball Bureau, 2026), disjoint roles and complete eligible stints. It reports both totals and exposure-based comparisons because shared playing time can increase total-WAR agreement. Forty-seven unresolved optional bulk-player decompositions are excluded from this supplement while independently reconciled supplied team Totals remain eligible. These comparisons describe agreement with bWAR; individual performance has no observed ground-truth WAR label.

### 2.7 Learned team-wins comparison

A bounded additional study compared a two-hidden-layer neural ensemble with ridge regression, histogram gradient boosting and saved scalar/run baselines. CORE contained team scoring, prevention, event rates and league context; replay also included the origin-season record. AUGMENTED added six conventional iWAR component rates. CORE already included observed runs scored and allowed, so this tests incremental information conditional on strong run-based inputs. Each family and feature set received its own pre-1970 development selection. Annual evaluation then fitted only target years earlier than the scored year, for both reconstruction and replay. This differs from the retrospective leave-one-year-out reconstruction calibration above. The primary comparison retained the development-selected Ridge AUGMENTED comparator and original 1,552 replay pairs across 1970–2025. Three fixed-seed neural fits formed each annual ensemble.

The study was amended after initial evaluation exposed a numerical defect. An exact-zero constant detector treated floating-point residuals in a mathematically constant position feature as meaningful variation, producing unstable standardized inputs when the designated hitter introduced real variation. A generic, training-only machine-precision constant guard repaired that implementation. The original outputs remain archived. Corrected development selection was frozen before corrected annual fitting, but after inspection of the initial evaluation; these results are amended historical evidence, rather than an untouched holdout. Supplement S9 provides the chronology, feature definitions, settings and preserved before/after evidence.

Publication-stage descriptive tables match saved evaluation rows across learned and provider methods without changing the primary cohort. Learned models retain full earlier internal history; saved provider calibrations retain their own matched histories. An evaluation-row match does not equalize training. Existing conditional intervals are used only for exact matching contrasts. No new fits, simulations or resampling were performed for publication integration.

### 2.8 Research procedures and software assistance

OpenAI ChatGPT and Codex assisted with study design, software implementation, analysis specifications, literature searches and manuscript preparation. The author directed the work and is responsible for the study decisions, verification of the reported evidence and final text. The retained protocols and review records identify the completed calculations and the scope of the checks performed. This study used historical professional baseball records and published statistical summaries. Research ethics approval and informed consent were not applicable; no participants were recruited and no private participant information was collected.

# 3 Results

## 3.1 Closer same-season reconstruction

Table 1 shows the primary common-sample comparison. iWAR RA had the lowest reconstruction error among the four WAR measures. Its MAE was 3.695 wins per 162, compared with 4.724 for bWAR and 4.744 for fWAR. Using the full-precision losses, these are reductions of 21.793% and 22.112%, respectively. The improvement is approximately one win per 162 in the average absolute error. These percentages describe proportional reductions in reconstruction error. Game-by-game classification accuracy is a separate quantity.

**Table 1. Fixed-map performance on identical common-cohort observations.** Reconstruction uses 1,471 team-seasons in 79 years; replay uses 1,441 consecutive franchise pairs in 78 target years. Errors are wins per 162 played games. Lower values indicate smaller error. RA denotes runs-allowed pitching; FIP denotes fielding-independent pitching; MAE denotes mean absolute error; RMSE denotes root mean squared error. All WAR methods use the fixed map $0.294+\mathrm{WAR}/G$.

| Method | Reconstruction MAE | Reconstruction RMSE | Replay MAE | Replay RMSE |
|---|---|---|---|---|
| iWAR RA | 3.695 | 4.654 | 9.150 | 11.550 |
| iWAR FIP | 3.958 | 4.976 | 8.839 | 11.075 |
| bWAR | 4.724 | 5.864 | 9.445 | 11.727 |
| fWAR | 4.744 | 5.941 | 8.882 | 11.076 |
| PythagenPat | 3.263 | 4.122 | 8.924 | 11.310 |
| Pythagorean exponent 2 | 3.381 | 4.287 | 9.244 | 11.731 |
| .500 | 11.300 | 13.928 | 11.249 | 13.872 |
| Previous record | — | — | 9.573 | 12.027 |

The paired RA-minus-bWAR MAE difference was −1.030, with conditional 95% interval [−1.367, −0.706]. The corresponding difference against fWAR was −1.049 [−1.398, −0.725]. RMSE also favored RA: 4.654, compared with 5.864 for bWAR and 5.941 for fWAR. The reconstruction advantage therefore appears under both average absolute error and a measure that gives more weight to large misses.

The FIP alternative also improved reconstruction over the two providers, although its error exceeded that of RA. Its MAE differences were −0.767 [−0.973, −0.561] against bWAR and −0.786 [−0.973, −0.600] against fWAR. The two pitching conventions consequently share a reconstruction improvement over the providers. Both alternatives also retain common batting, running and other accounting choices, so the aggregate comparisons leave their separate contributions unresolved.

PythagenPat had lower reconstruction error than every WAR measure, with MAE 3.263 and RMSE 4.122. The exponent-two Pythagorean reference also performed well. These comparisons establish the scale of the remaining error: observed team runs contain a strong relationship with the completed team record, and converting the more elaborate WAR totals does not improve on the best run-based reference in this sample. The result leaves a clear advantage for iWAR RA within the four-WAR comparison while identifying an even stronger team-record comparator.

### 3.2 Replay and the effect of conversion

Moving the target to the following season changed the ordering. Under the fixed map, iWAR RA improved on bWAR by 0.295 MAE wins per 162, with paired interval [−0.496, −0.105], but had greater error than fWAR by 0.268 [+0.072, +0.475]. iWAR FIP performed better than bWAR by 0.606 [−0.825, −0.382] and was close to fWAR. The FIP-minus-fWAR MAE difference was −0.043 [−0.167, +0.087], while the RMSE difference was approximately −0.000429. The paired display-rounding bound for the MAE comparison was ±0.104. These data leave the fixed-replay FIP–fWAR difference unresolved.

Calibrating each measure provided a way to examine whether the fixed conversion drove the reconstruction results. On the original 1,471 reconstruction rows, calibrated MAE was 3.682 for RA, 3.952 for FIP, 4.709 for bWAR and 4.707 for fWAR. The RA differences remained −1.027 [−1.336, −0.746] against bWAR and −1.025 [−1.284, −0.771] against fWAR. Their similarity to the fixed differences indicates that an intercept and slope account for little of the pooled reconstruction advantage. The FIP differences likewise remained close to their fixed counterparts.

Replay was more sensitive to conversion. Table 2 reports the calibrated predictors and unchanged .500 reference on the same 1,282 pairs. On this subset, the fixed RA-minus-fWAR MAE difference was +0.231 [+0.024, +0.452]. After calibration, it became −0.148 [−0.295, −0.008]. FIP-minus-fWAR changed from fixed −0.067 [−0.194, +0.071] to calibrated −0.176 [−0.263, −0.088]. The change in sign for RA occurs on identical evaluation rows and therefore reflects the estimated conversion. Every method is scored after the same training-year exclusions.

**Table 2. Common-cohort replay after chronological scalar calibration.** Errors are wins per 162 played games. MAE denotes mean absolute error; RMSE denotes root mean squared error. All predictors use the same 1,282 pairs in 67 target years. Each scored year uses a conversion estimated from strictly earlier target years; the .500 reference remains fixed. The run-based comparisons are point estimates; paired intervals were calculated for the prespecified iWAR–provider contrasts.

| Calibrated predictor | MAE | RMSE |
| --- | --- | --- |
| Linear run differential | 8.177 | 10.169 |
| PythagenPat | 8.198 | 10.165 |
| Pythagorean exponent 2 | 8.223 | 10.180 |
| iWAR FIP | 8.260 | 10.241 |
| iWAR RA | 8.288 | 10.263 |
| fWAR | 8.436 | 10.461 |
| bWAR | 8.838 | 10.796 |
| Previous record | 8.520 | 10.547 |

| Calibrated predictor | MAE | RMSE |
|---|---|---|
| .500 reference, unchanged | 10.993 | 13.527 |

The run-based predictors also benefited from estimated conversion. Calibrated run differential had the lowest MAE, and calibrated PythagenPat had the lowest RMSE. Both had lower point errors than all four calibrated WAR measures. Their differences from iWAR FIP were small, but they are relevant to the practical question of which season summary carries information about the following year. The result also prevents the calibration analysis from being read as a competition restricted to the most favorable WAR ordering.

The fixed and calibrated effects use identical evaluation rows. The contrast between tasks is substantial: the reconstruction advantage persists with little change, whereas replay comparisons depend more strongly on the mapping. Calibration changes the use made of a team summary. It leaves the underlying player allocations untouched.

### 3.3 Variation across eras and source coverage

The pooled reconstruction advantage was uneven across historical periods. Table 3 shows the fixed RA MAE differences. The earliest period contributed 563 rows and relatively large differences from both providers. The contemporary common sample contributed 240 rows from eight observed years, with much smaller differences. The positive reconstruction result is therefore most pronounced in the earlier history represented by this sample.

**Table 3. Same-season fixed-map MAE differences by reporting period.** Differences are iWAR RA minus provider, in wins per 162; negative values favor RA. Each period has fewer than the ten nonempty five-year blocks required for the primary interval calculation, so the table reports descriptive point estimates.

| Reporting period | Team-seasons | Five-year blocks | RA−bWAR MAE | RA−fWAR MAE |
|---|---|---|---|---|
| 1908–1945 | 563 | 8 | −1.525 | −1.632 |

| Reporting period | Team-seasons | Five-year blocks | RA−bWAR MAE | RA−fWAR MAE |
|---|---|---|---|---|
| 1946–1968 | 374 | 5 | −1.017 | −0.785 |
| 1969–1993 | 182 | 3 | −0.750 | −1.195 |
| 1994–2014 | 112 | 1 | −1.117 | −0.913 |
| 2015–2025 | 240 | 3 | −0.058 | −0.046 |

The contemporary differences illustrate why both error measures and source precision matter. The RA MAE advantage against bWAR was 0.058 wins per 162, slightly larger than its ±0.050 display bound. Its advantage against fWAR was 0.046, within the ±0.100 bound. In RMSE, bWAR was better than RA by 0.078; RA was better than fWAR by 0.061. After calibration, the contemporary RA MAE differences were −0.181 against bWAR and −0.352 against fWAR, while the corresponding RMSE differences were −0.130 and −0.427. The pooled findings therefore do not imply one unchanging ordering across periods and mappings.

Each era's reconstruction MAE point difference favored RA under both mappings. The available block counts limit how far that pattern can be taken. A pooled interval combines the observed history; it does not provide five separately resolved era conclusions. Supplementary tables retain supported one-year-block sensitivity intervals, together with the primary five-year rule and all annual values. Figure 1 shows the missing-year gaps directly. Figure 2 compares the fixed and calibrated effects on matched evaluation rows.

The expansion of the original benchmark was also concentrated historically. Its 69 newly compared origin or same-season years consist of 59 years before 1968, excluding the original 1927 example, plus 1988–1997. No post-1997 origin year was newly added. The exact 279-row overlap with the earlier reconstruction comparison retained RA MAE/RMSE of 3.770/4.802, bWAR 4.018/4.989 and fWAR 3.965/5.002 under the present per-162 scoring. The 1,192 added rows had RA 3.677/4.619, bWAR 4.890/6.051 and fWAR 4.926/6.140. The larger pooled

advantage arose mainly from adding different historical observations while preserving the original overlap.

Replay periods give a further view of the differences. Fixed RA-minus-fWAR MAE was positive in all five target-era groups, ranging from +0.073 to +0.531. Fixed FIP-minus-fWAR changed sign across periods. Calibration changed these patterns again: RA-minus-fWAR became negative in the first two groups and positive in the last three. These patterns are descriptive because the era groups contain only two to eight five-year blocks. Their value is to show where the pooled comparison changes, while retaining the available sample behind each result.

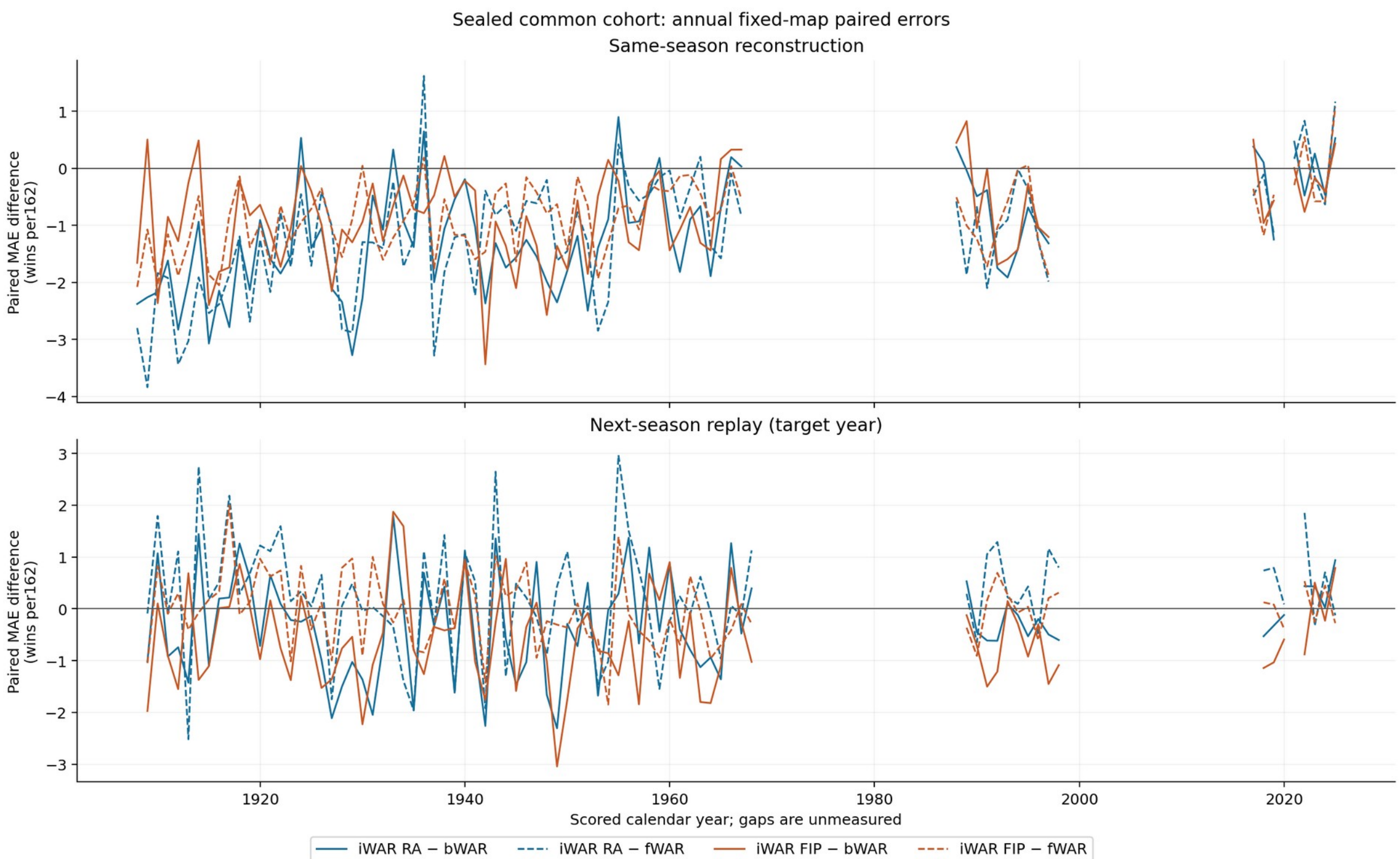

Figure 1. Annual fixed-map paired MAE differences on the common cohort. Gaps identify years without a common comparison. Negative differences favor the corresponding iWAR measure.

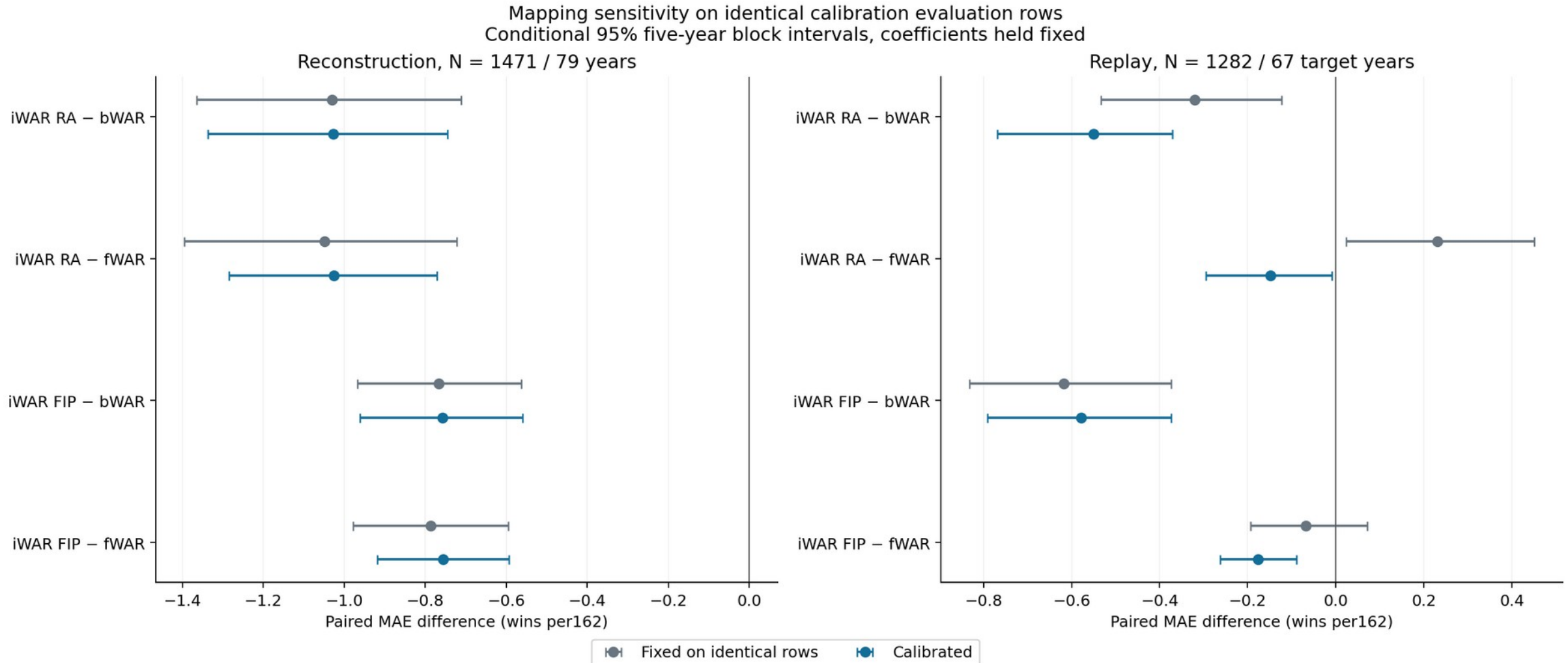

Figure 2. Fixed and calibrated paired effects on identical common-cohort evaluation rows. Same-season and replay results retain their distinct training rules and sample sizes.

### 3.4 The broader bWAR comparison

The secondary comparison added history that the common fWAR requirement excluded. Its 2,554 reconstruction rows contain the 1,471 original common rows, 1,068 additional rows from the 39 fWAR-missing origin years, and 15 otherwise eligible supplied-bWAR rows rejected by the original four-method record or scope requirements. The replay partitions contain 1,441, 1,068 and 15 pairs, respectively. The added observations remain separate in the supplementary results so that their contribution to the broader estimate can be examined.

Table 4 shows that the iWAR advantage over bWAR persisted in the expanded reconstruction, but its magnitude decreased. RA MAE was 3.755, compared with 4.452 for bWAR, a difference of −0.697 [−0.990, −0.443]. FIP MAE was 3.891, a difference of −0.561 [−0.772, −0.346]. Thus the common-sample result was not solely a product of excluding every fWAR-missing year, although those exclusions materially affected the size of the comparison.

**Table 4. Secondary fixed-map comparison with bWAR.** Differences are iWAR minus bWAR, in wins per 162. "Missing-fWAR" identifies added observations classified by origin-year availability. Intervals are conditional five-year calendar-block summaries. An interval is withheld when fewer than ten nonempty blocks are available.

| Task and population | N / years | RA−bWAR MAE [95% interval] | FIP−bWAR MAE [95% interval] | Blocks |
|---|---|---|---|---|
| Reconstruction, full secondary sample | 2,554 / 118 | −0.697 [−0.990, −0.443] | −0.561 [−0.772, −0.346] | 24 |
| Reconstruction, added missing-fWAR years | 1,068 / 39 | −0.244 [withheld] | −0.284 [withheld] | 9 |
| Replay, full secondary sample | 2,524 / 117 | −0.197 [−0.336, −0.073] | −0.541 [−0.718, −0.369] | 24 |
| Replay, added missing-fWAR years | 1,068 / 39 | −0.077 [−0.224, +0.050] | −0.461 [−0.734, −0.237] | 10 |

The added missing-fWAR history had smaller reconstruction differences than the common sample: −0.244 for RA and −0.284 for FIP. Its 39 years occupied nine nonempty five-year blocks, so reconstruction intervals were withheld under the unchanged rule. For replay, the RA difference in those added years was −0.077 [−0.224, +0.050], with an RMSE difference of +0.007. FIP retained a clearer MAE advantage against bWAR. The small final partition of 15 replay pairs favored bWAR over RA by 0.738 MAE wins per 162 and remained included in the broader summary.

The full secondary replay had a negative RA MAE interval, but its RMSE interval crossed zero: −0.095 [−0.201, +0.003]. This is another instance in which the chosen loss affects the conclusion. Broader contemporary coverage also changed the RA-minus-bWAR fixed reconstruction RMSE difference from +0.078 in the 240-row common sample to −0.242 in the 330-row secondary sample. Both results are retained, with their different memberships explicit. Calibration preserved the broader reconstruction advantage. The RA and FIP MAE differences against bWAR were −0.716 [−0.985, −0.463] and −0.560 [−0.766, −0.340]. The broader calibrated replay contained 2,359 pairs in 106 target years, with corresponding differences −0.377 [−0.532, −0.248] and −0.459 [−0.625, −0.310]. Run-based predictors and iWAR FIP were closely grouped on these same rows: run differential MAE/RMSE was 8.051/10.002, PythagenPat 8.068/10.005, and FIP 8.057/10.001. Their small point ordering depended on the error measure.

Figure 3 compares the common and broader bWAR evidence. The expansion shows how much of a conclusion can depend on the historical intersection available to all methods. Equal-season weighting can change the influence assigned to observed years, but it cannot supply the missing fWAR observations. The differences between the samples combine era, eligibility and source availability; they do not isolate a causal effect of provider access.

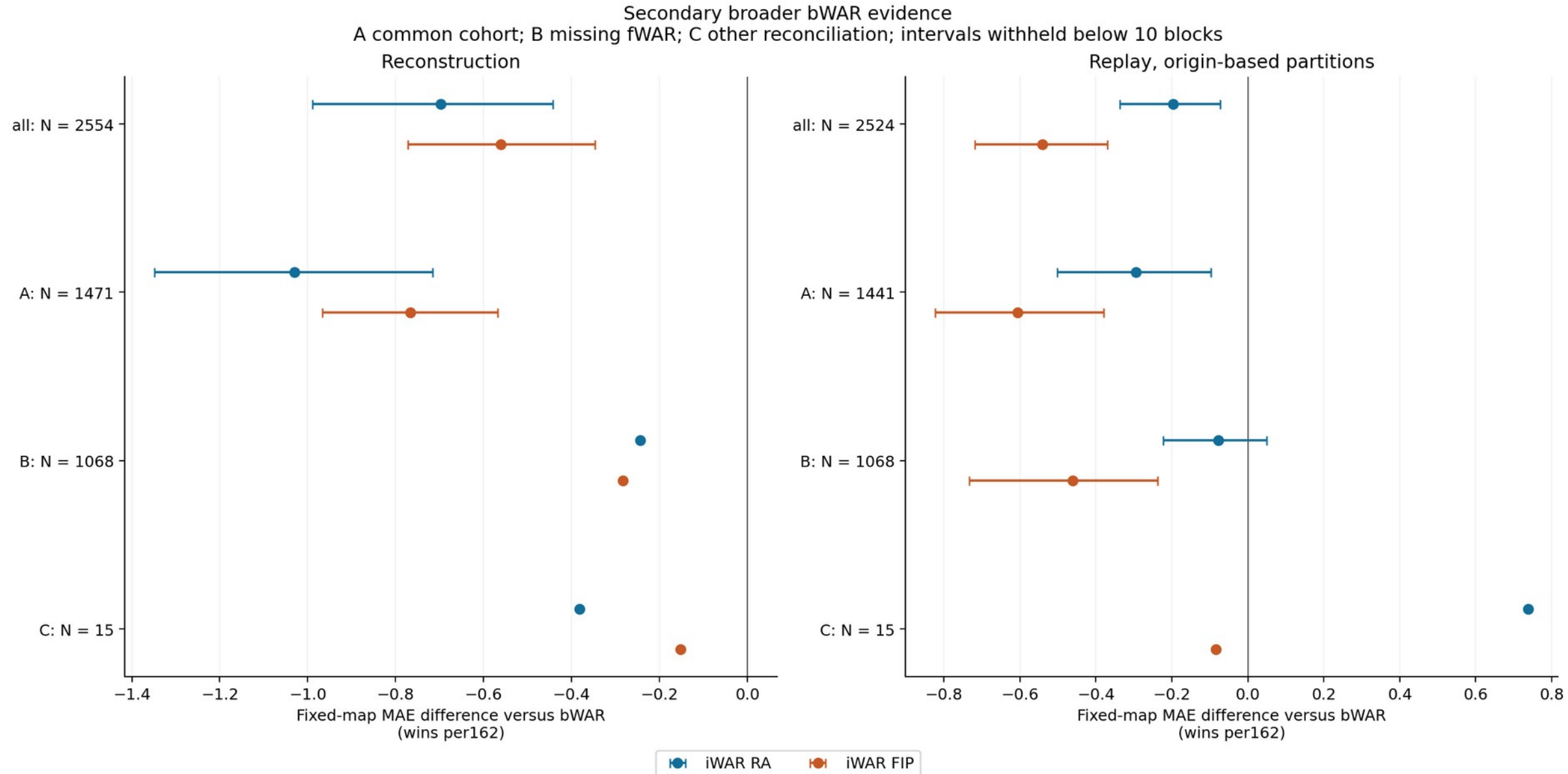

Figure 3. Common and broader bWAR comparisons. Secondary replay partitions follow origin-year source availability. Intervals appear only where the specified block-support requirement is met.

### 3.5 What the accounting comparison reveals

PythagenPat's same-season MSE was 16.995 squared wins per 162, compared with 21.658 for RA, 24.759 for FIP, 34.390 for bWAR and 35.300 for fWAR. For RA, the excess MSE of 4.663544 separated algebraically into a squared prediction distance of 7.050269 and an error cross-term of −2.386725. The corresponding squared distances were 11.568 for FIP, 21.210 for bWAR and 22.046 for fWAR. Every method had a negative cross-term, which partly offset its distance from the run-based prediction.

This description helps locate the empirical advantage. The RA prediction remains closer to the observed-run expectation, and its smaller squared distance contributes to its lower reconstruction MSE. The common conversion calibration removes little of the gap. A complete aligned provider component ledger was unavailable, so the remaining difference cannot be apportioned causally among parks, replacement, batting or defense. The supplement reports the full arithmetic, weighting sensitivities and era results.

League centering offers a separate explanation for the small pooled signed bias. On complete paired league scope, the combination of centered runs and the replacement allowance fixes the total implied wins. An incomplete common intersection and equal-row per-162 weighting need not satisfy that identity exactly. Small pooled bias therefore should be read together with MAE and RMSE, which assess how the implied wins are distributed among teams.

### 3.6 Agreement at the player level

The bWAR supplement contains 90,891 combined player-team observations and 83,642 complete union player-seasons across 118 years. Combined RA correlation with bWAR was 0.934, with MAE 0.406 wins; FIP correlation was 0.916, with MAE 0.449. Pitching-only correlations were 0.902 for RA and 0.854 for FIP across 48,136 rows. A separate 29,518-row pitching-rate comparison gave correlations of 0.800 and 0.674. Because both the population and units change, those total-to-rate differences cannot quantify the effect of playing time by themselves.

An aligned 36,355-row arithmetic comparison showed a mean RA total difference of +0.050 wins, comprising +0.047 pitching credit and +0.003 remaining-role credit. FIP showed +0.047 in total, comprising +0.044 and +0.003. The pooled components share the same sign. Individual observations can still contain offsetting differences, and the remaining-role quantity combines several contributions. These results document broad agreement with an established valuation while leaving the correctness of individual assignments open.

### 3.7 Learned team-wins comparison

The neural ensemble had larger next-season RMSE than the development-selected Ridge AUGMENTED comparator: 10.163 versus 9.946 credited wins per 162 on the original 1,552 pairs (Table 5). The difference was +0.217, with conditional interval [0.015, 0.474]. Ridge also had smaller reconstruction error. The modern point differences favored Ridge, but intervals were withheld because only four nonempty calendar blocks were available. These results concern the tested inputs, architectures and development procedure.

**Table 5. Amended learned-model comparisons.** Equal team-season weights; unbounded predictions; RMSE in credited wins per 162. Positive differences mean larger neural error. The full-period intervals are saved conditional pointwise 95% five-year-block intervals with predictions fixed. Modern intervals are unavailable under the ten-block rule.

| Target and period | N | Neural AUG | Ridge AUG | Difference | 95% interval |
|---|---|---|---|---|---|
| Next season, 1970–2025 | 1552 | 10.163 | 9.946 | +0.217 | [0.015, 0.474] |
| Next season, 2010–2025 | 480 | 10.900 | 10.462 | +0.438 | Unavailable |
| Same season, 1970–2025 | 1558 | 4.282 | 3.848 | +0.434 | [0.340, 0.533] |
| Same season, 2010–2025 | 480 | 4.368 | 3.932 | +0.436 | Unavailable |

Adding components increased neural RMSE by +0.165 in replay and +0.029 in reconstruction; both saved intervals included zero. The corresponding replay point differences were +0.012 for Ridge and +0.019 for boosting. Reconstruction differed: Ridge improved by 0.089, while boosting worsened by 0.315. Those newly summarized Ridge and boosting contrasts have no saved intervals. Because feature sets were selected separately, these are effects under the specified selection procedure, rather than fixed-hyperparameter ablations.

The provider-matched replay subset contained 504 pairs across 18 target years. On those identical rows, fixed-map fWAR RMSE was 11.291, iWAR FIP 11.287, Ridge AUGMENTED 10.316 and neural AUGMENTED 10.297. The neural–Ridge point ordering therefore reversed on this smaller descriptive subset. Compatible v0.23 fWAR predictions were verified on these same 504 pairs; its full-cohort RMSE of 11.076 on 1,441 pairs is contextual and cannot be ranked against the new full-cohort Ridge result. No broad-cohort interval is transferred to this subset. Supplement S9 reports fixed and calibrated provider comparisons and the larger bWAR overlap.

## 4 Discussion

The strongest result is the same-season comparison. iWAR RA reduced reconstruction MAE by approximately 22% relative to both provider WARs on the matched 79-year sample, and its advantage appeared under RMSE as well. Estimating each method's intercept and slope preserved almost all of the MAE difference. The broader bWAR analysis retained a smaller advantage across 118 years. These findings make the reconstruction superiority specific and measurable: it applies to the completed team records, the observed historical inputs and the stated conversions.

The estimated-conversion analysis is useful because WAR is already presented in win units. That presentation can make the conventional linear relationship seem like the only reasonable way to evaluate a team total. Allowing a scalar conversion asks a practical additional question: does the difference survive when each measure receives an appropriate overall level and slope? For reconstruction, the answer was largely yes. For replay, the change was substantial enough to reverse the RA–fWAR ordering on identical rows. Reporting only one of those mappings would conceal information about the measures' behavior.

The run-based comparisons sharpen the interpretation. PythagenPat reconstructed completed records more accurately than every WAR method, and calibrated run differential and PythagenPat produced the lowest point errors in the common replay sample. A reader interested solely in team-record prediction would reasonably consider those simpler summaries. WAR serves an additional purpose by distributing value among players and aspects of play. That purpose creates accounting decisions whose consequences can be examined at the team level, while their individual validity requires other evidence.

The distinction is exact for a conserving pitching–fielding transfer. In the RA implementation, an additional run assigned to fielding is subtracted from pitching credit, with corresponding centering. The team total is unchanged. Any team-win loss therefore gives the same answer for all such allocations. This is a direct limit on what improved team reconstruction can establish. Richer defensive observations, carefully chosen reference players and studies with known simulated responsibility can address different parts of the attribution problem. Recent work on defensive measurement also emphasizes that improving outcome prediction can alter the variation assigned to individual responsibility (Judge, 2025).

Historical breadth is a substantive contribution of this study. The older and newer common observations gave different effect sizes, and the added bWAR history reduced the pooled gap further. These results show why historical availability belongs in the main findings. The 39 fWAR-missing years are not an interchangeable set of random omissions. They occupy identifiable stretches of baseball history, and the methods need not behave identically in those stretches. The original-to-expanded comparison provides direct evidence that adding these kinds of observations can change the apparent scale of superiority.

The literature also helps define the contribution. openWAR established a public-data approach to overall player valuation, while Grid WAR investigated how game-level nonlinear value affects starting pitchers. The present work adds a common historical evaluation that follows the same methods across alternative targets, conversions and available populations. Related Retrosheet work continues to emerge, including Brown's (2026) mWAR preprint. Its abstract describes another skill-oriented alternative, but its full procedures were unavailable for the present comparison. We therefore make no numerical performance claim against that method or other alternatives lacking comparable retained predictions.

The learned-model analysis adds a bounded test of model complexity. Its designated comparison favors ridge regression, and the neural component intervals leave both modest benefit and harm compatible with the retained evidence. Reconstruction's Ridge component improvement also prevents a general claim that components have no value. CORE already contains realized scoring and prevention. Conditional predictive value in that setting does not measure the accuracy of individual batting, running or fielding credit. Amended selection, historical input revisions and limited modern blocks further constrain generalization. The neural results belong alongside the historical comparison as an additional analysis; they do not replace its conventional valuation or justify a new individual WAR definition.

Several limitations affect the conclusions. The methods were developed before this expanded benchmark, and some seasons had already been examined during development. The protocol was frozen before the new comparisons, but project-wide independence cannot be claimed. The broader bWAR analysis was informed by the primary findings. Its separate specification and preserved comparison rows clarify the sequence of decisions without removing that prior knowledge.

Replay evaluates persistence in franchise summaries. Roster turnover, aging, injuries and player-level projections are absent, and current-vintage provider WAR may include later revisions. The chronological scalar fit prevents later target outcomes from entering that fit, but it does not reconstruct the information available to an analyst at each historical date. A future as-of-date study would need archived inputs and a clearly defined forecast date. The present results support conclusions about retrospective franchise replay under the retained inputs.

The uncertainty calculations also have a limited scope. Resampling calendar blocks preserves local historical groupings while holding estimates and coefficients fixed. Dependence can cross block boundaries, and baseball's run environments and rules change over time. The reported intervals describe the specified resampling procedure; they do not encompass every source of uncertainty. Era groups are particularly limited, with insufficient five-year blocks for their primary intervals. Provider rounding further matters when the observed differences are only a few hundredths of a win.

The public package contains the actual estimator, provider adapters, benchmark and neural-study implementation, including the original and corrected constant handling. It also reproduces publication reporting offline from approved aggregates. Source-level reruns require compatible Retrosheet inputs, provider observations and retained membership records under their applicable terms. Public code availability does not confer redistribution rights for those observations.

## 5 Conclusion

iWAR RA reconstructs historical team records more closely than bWAR and fWAR on the matched common sample, with an approximately 22% reduction in average absolute error. Most of that advantage persists after estimating a separate conversion for each method. A broader bWAR comparison retains the finding at a smaller magnitude and shows that historical coverage changes the estimated effect.

Next-season replay gives a different result. Fixed-map iWAR FIP and fWAR have nearly identical errors, while calibration changes several pairwise rankings. Run-based summaries remain strong comparators. A credible assessment of WAR should therefore name the target, report the conversion and show which historical observations support the conclusion. These comparisons establish where iWAR improves team-level accounting and provide a reproducible set of aggregate results for evaluating that improvement. They leave the allocation of individual credit as a separate empirical question.

## Data and code availability

The scientific code, configurations, protocols, correction records and aggregate publication evidence are available at https://github.com/MRDOANE/iwar-historical-benchmark. Version v1.0.0 is archived at https://doi.org/10.5281/zenodo.23455100 and corresponds to commit `19d0add57222d54f22a9fd40166153213254776c`. The release includes the conventional valuation specification, selected existing tests and environment declarations. Its offline reporting mode rebuilds publication tables and figure sources from released aggregates without fitting or network access. Its scientific mode requires the genuine private inputs, documented dependencies and original computational gates; that mode was not executed during publication preparation.

Raw provider observations, provider-valued team and player rows, calibration memberships, exact row predictions, source crosswalk extracts, model objects and resampling draws are excluded from the public deposit. Their local inventories and fingerprints document provenance. Full independent recomputation requires authorized access to those inputs; the released code is not a replacement for missing source data. The package distinguishes original MIT-licensed code and documentation from third-party data, dependency and source terms.

The Chadwick Persons Register crosswalk used commit `781c175e748b89e5730fd3f5de0086b9355b2f92`, dated 2 October 2026, under the Open Data Commons Attribution License 1.0. Source-data and software-dependency terms remain separate from the license for original code. The source manifest records cached source vintages; this historical evaluation does not reconstruct archived real-time provider releases.

## Acknowledgments

The code and aggregate evidence supporting this work are available at https://github.com/MRDOANE/iwar-historical-benchmark.

The information used here was obtained free of charge from and is copyrighted by Retrosheet. Interested parties may contact Retrosheet at www.retrosheet.org.

# Supplementary material for Comparing wins above replacement across baseball history

Michael Doane

Independent Researcher, Cary, North Carolina, USA

## S1. Purpose and source records

This supplement gives the conventional iWAR definitions, the rules used to assemble the comparison samples, and additional results needed to interpret the team-level findings. The main article compares two questions: how closely a team's WAR total reconstructs its credited wins in the same season, and how well its current-season summary carries forward to the next season. The detailed definitions below explain where observed scoring enters those calculations and which aspects of player credit remain unresolved.

The valuation uses Retrosheet regular-season event and statistical records (Retrosheet, 2026). The historical inventory contains 2,594 American League and National League team-seasons from 1908 through 2025. Of these, 2,568 have complete conventional iWAR products. Each event and player summary is assigned to the same certified game inventory before calculation. A complete product means that the core identities, roles, exposures and charged totals required by the stated convention are available. It does not mean that every physical skill has been observed. Missing core inputs propagate to the dependent total; an incremental skill assigned an assumed-average contribution of zero retains an unavailable skill measurement.

The Retrosheet source version used for the expansion records an August 9, 2026 site update, a December 7, 2025 game-log index update, and a terminal season of 2025. Provider comparisons use the actual Baseball-Reference and FanGraphs values obtained or preserved during the study. Those values have mixed retrieval dates and may include retrospective revisions. The study therefore evaluates the versions available when the comparison was assembled. It does not reconstruct every statistic as it would have appeared at the end of its original season.

The conventions were developed on ten seasons: 1927, 1975, 2017–2019 and 2021–2025. The expanded comparison follows frozen rules applied to additional historical years. Its new years are new to this comparison; they are not an untouched test of every prior project decision. The broader Baseball-Reference analysis was designed after the common-cohort results were known, with its rules frozen before the additional losses were calculated.

## S2. Conventional iWAR calculation

### S2.1. Components and units

For player $i$ within a league-season stint, let $\mathrm{BAT}_i$, $\mathrm{RUN}_i$, $\mathrm{FIELD}_i$, $\mathrm{PITCH}_i$, and $\mathrm{POS}_i$ denote batting, running, fielding, pitching, and position contributions in runs. Let $\mathrm{CENTER}_i$ denote the sum of the applicable league-role centering terms. Writing this total run contribution as $R_{i,\mathrm{RA}}$, the runs-allowed version is

$$R_{i,\mathrm{RA}} = \mathrm{BAT}_i + \mathrm{RUN}_i + \mathrm{FIELD}_i + \mathrm{PITCH}_{i,\mathrm{RA}} + \mathrm{POS}_i + \mathrm{CENTER}_{i,\mathrm{RA}},$$

$$\mathrm{iWAR}_{i,\mathrm{RA}} = \frac{R_{i,\mathrm{RA}}}{\mathrm{RPW}} + \mathrm{REPL}_{i,\mathrm{ordinary}} + \mathrm{REPL}_{i,\mathrm{pitching}}. \quad \text{(S1)}$$

Here RPW is the league-season conversion from runs to wins. Replacement contributions are already expressed in wins. The FIP version changes the pitching component and the centering term associated with that pitching convention. Cross-league stints are converted within their own league-seasons and then added in wins.

Actual event roles determine whether batting belongs to ordinary batting or pitcher batting. Batting by fielders, designated hitters, pinch hitters and pinch runners belongs to the ordinary pool. Pitcher-role batting is kept separate. A player's career label does not determine an event's role. This distinction matters for two-way players, substitutions and replacement credit: pitcher-role batting receives no second ordinary replacement allowance.

S2.2. Run expectancy and standardized batting

A base-out state $s$ records the occupied bases and zero, one or two outs, giving 24 states. League-season run expectancy is estimated from completed three-out half-innings. Each state has a twenty-observation prior toward the pooled empty-base reference multiplied by $(3-o)/3$, where $o$ is the number of outs. A state without observations uses that finite pooled reference. The treatment of automatic runners is described in Section S2.5.

For an event, its recorded run value $V$ is the adjusted runs scored plus the run expectancy after the event minus the run expectancy before it. A third out sets the after-state expectancy to zero. A walkoff retains the ordinary remaining-state expectancy; the fact that the game ended does not erase that state from this valuation convention.

The batting categories distinguish singles, doubles, triples, ordinary and explicitly identified inside-the-park home runs, walks, intentional walks, hit batters, strikeouts and strikeouts on which the batter reaches, sacrifices, errors, fielder's choices, interference, multiple outs and other outcomes. A plate appearance accompanied by a steal, caught stealing, pickoff, wild pitch, passed ball, balk or other-advance flag receives a compound category. These categories describe recorded outcomes.

Each category has a specified normal transition $B(c,s)$. Hits move runners station to station. Walks, hit batters, interference, errors and strikeout-reached outcomes use forced chains. A fielder's choice replaces the first-base incumbent with the batter and adds an out when that incumbent exists. An ordinary out or sacrifice retires the batter while holding incumbents. A double play retires the batter and the runner on the lowest occupied base; a triple play clears the bases; a home run scores everyone. Third outs clear the state. The expected advancement associated with sacrifice outcomes enters through the estimated category advancement below.

Define advancement beyond the normal transition as

$$A = V - B(c,s). \quad \text{(S2)}$$

Its league-season mean $\overline{A}(c,s)$ uses twenty-event shrinkage toward the category-wide advancement mean. The estimated category-state value and standardized category weight are

$$m(c,s)=B(c,s)+\overline{A}(c,s),$$
$$w(c)=\sum_s q_s m(c,s), \quad \text{(S3)}$$

where $q_s$ is the distribution of pre-plate-appearance states among all official plate appearances in the league-season. The same $q_s$ is used for every category. Thus two occurrences of the same batting category receive the same weight within that reference league-season, even if the particular events occurred in different base-out states. The player's raw batting contribution sums those weights after subtraction of the mean weight for the actual batting-role population.

The calculation standardizes event timing while retaining the observed season's run and advancement references. Sparse pitcher-batting references use a 100-plate-appearance prior toward the covered same-season pitcher-batting pool. Empty or unresolvable role pools remain explicitly recorded. Batting uses a neutral component park factor.

### S2.3. Running

Running credit uses the same advancement reference as batting. A discretionary advancement opportunity is eligible only when the event category is supported, ownership is known and every relevant endpoint can be resolved from post-event bases, scored runners or recorded retirements. The supported clean plate-appearance categories are singles, doubles, triples, sacrifice hits, sacrifice flies and other outs; explicitly recorded batter extra-base or out attempts may also participate. Post-win discretionary advancement is unavailable.

For an eligible event, the running surplus is

$$\text{surplus} = A_{\text{actual}} - \overline{A}(c,s). \quad \text{(S4)}$$

Equal shares are assigned across its certified opportunities. This is an explicit responsibility convention. It can credit a normal advancer when another runner made the additional advance, so an individual share should not be read as a direct estimate of speed or decision quality. Unknown or unrecorded opportunities are not treated as observed decisions to hold a runner.

The batting and running pieces use one reference: $w(c)+\text{surplus}=V_{\text{actual}}-m(c,s)+w(c)$. For example, a batter's single receives standardized average advancement, and the eligible running opportunities receive the departure from that same average. Automatic advancement on home runs has zero discretionary running value. Errors, fielder's choices, strikeouts, multiple outs, compound events and uncertified opportunities retain an assumed-average incremental running contribution. Isolated, unambiguous non-plate-appearance steals, caught stealing and pickoffs receive their transition value once and are absent from the plate-appearance batting reference. Running also uses a neutral component park factor.

### S2.4. Coarse fielding and defensive support

The conventional fielding component uses recorded primary conversions and assigned nonconversions. Its inputs support a coarse allocation of run credit. They do not measure range, arm strength, framing or the difficulty of every opportunity in the manner of a tracking-based defensive metric.

Eligible balls in play exclude strikeouts and ordinary home runs and include explicitly identified inside-the-park home runs. Active positions 1–9 participate. Each contact supplies one primary conversion or one total assigned nonconversion. A hit followed by a runner out remains a primary nonconversion. A successful initiating retirement counts once on a double play, triple play or successful fielder's choice. An error after an already successful conversion does not add a second full nonconversion.

Known conversion credit belongs to the initiating fielding sequence. A primary error uses the responsible position, and a failed fielder's choice uses its initiator where identified. The eventual retriever of a hit is not automatically treated as the defender who failed to convert it. Unknown responsibility is allocated fractionally and flagged. A missing active player identity remains a missing core input.

Fractional responsibility uses same-season out-position profiles. For batter $b$, pitcher throwing hand $h$, and fielding position $p$, let $o_{bhp}$ be the number of identifiable primary outs. The estimated position share is

$$q_{bhp} = \frac{o_{bhp} + 100\, q_{\text{league, batterhand}, h, p}}{\sum_p o_{bhp} + 100}. \quad \text{(S5)}$$

Unknown or empty handedness cells use the specified pooled-hand references. This resembles the use of out-position information in historical defensive methods, with the present study's same-season prior and responsibility rules stated explicitly (Smith, n.d.).

For each player-position-throwing-hand stratum, let $O$ be primary conversions, $N$ assigned nonconversions, and $C = O + N$. If $r$ is the corresponding league conversion rate, raw outs above reference are $O - rC$. Run credit multiplies this quantity by the league-season standardized nonconversion-minus-conversion run gap. That gap uses the same all-plate-appearance state weights $q_s$ as batting, with the appropriate pitching-league run-expectancy reference. Fielding uses a neutral component park factor. Unmeasured extra outs, arm contributions, catching, blocking and framing remain within the stated residual or assumed-average conventions.

Every signed fielding contribution also becomes defensive support assigned to the active pitcher. Consequently,

$$\sum_i \text{FIELD}_i = \sum_j \text{support}_j,$$

$$\sum_j \text{PITCH}_{j,\text{RA}} + \sum_i \text{FIELD}_i = \sum_j \text{grossRA}_j. \quad \text{(S6)}$$

A pitcher's own fielding is subtracted as support and added to that player's fielding contribution once. Inherited-run responsibility and active-pitcher defensive support follow different allocation rules. Equation S6 conserves their combined team total. A change in fielding credit accompanied by the equal-and-opposite pitching change, including its corresponding centering, therefore leaves team RA WAR unchanged. Team-win reconstruction cannot determine which of those individual allocations is more accurate. The FIP convention below does not subtract defensive support.

S2.5. RA pitching, parks and automatic runners

Pitching workload uses literal integer outs, with innings $IP = \text{outs}/3$. Zero-out appearances retain their charged runs. Gross RA credit is expected runs minus charged runs:

$$\text{grossRA}_j = \sum_e \text{outs}_{je} \ \lambda_{\ell t} \ PF_e - \text{chargedRuns}_j^{\text{adj}},$$
$$\text{PITCH}_{j,\text{RA}} = \text{grossRA}_j - \text{support}_j, \quad \text{(S7)}$$

where $\lambda_{\ell t}$ is the covered league-season's adjusted runs per out and $PF_e$ is the run factor for the actual site. Both earned and unearned charged runs enter. Inherited runs retain official pitcher responsibility, and there is no second debit for inherited base traffic through run expectancy. Official strikeout and walk responsibility after partial-plate-appearance substitutions is distinguished from active-pitcher outs and support.

Automatic runners are resolved through the recorded placement and substitution history. If every placement in a season can be resolved, only those runners' own scored runs are removed from charged totals and the corresponding league run-rate and runs-per-win inputs. If that condition fails, all runs are retained consistently for the season and the convention is flagged. Other effects of beginning an inning with a runner remain part of the observed season.

The RA park factor compares both clubs' adjusted runs per recorded out at the site with road games of the site's home clubs. Its log rate ratio is multiplied by $n_{\text{home}}/(n_{\text{home}}+100)$, then normalized to a pitching-out-weighted league mean of one. Missing or nonpositive references receive a neutral factor. Actual sites and park changes are retained. This factor adjusts RA expectation once. Batting, fielding and FIP retain neutral component factors because the implementation does not estimate separate component-specific park effects. The same-season park estimate can reflect team, opponent and schedule differences.

S2.6. FIP pitching

The alternative pitching numerator is

$$N_j = 13HR_j + 3(BB_j + HBP_j) - 2K_j. \quad \text{(S8)}$$

Here home runs include explicitly identified inside-the-park home runs, and walks include intentional walks. The implementation has no infield-fly addition. With $RA9_{\ell t}$ denoting the adjusted league runs per nine innings, define

$$c_{\ell t} = RA9_{\ell t} - \frac{\sum_j N_j}{\sum_j IP_j},$$

$$\text{PITCH}_{j,\text{FIP}} = \frac{RA9_{\ell t} IP_j - N_j - c_{\ell t} IP_j}{9}. \quad \text{(S9)}$$

This expression is defined for a zero-out appearance. It uses the observed league run environment while assigning the individual pitching contribution through home runs, walks, hit batters and strikeouts. It receives no RA defensive-support subtraction, RA park factor, leverage adjustment, opponent-strength adjustment or starter/reliever adjustment. These choices differ from the full FanGraphs pitching WAR procedure (Slowinski, 2012); iWAR FIP is the study's declared alternative, with its own league centering.

S2.7. Position, centering and replacement

The position schedule assigns the following run values per 1,458 defensive innings: catcher +12.5; first base −12.5; second base, third base and center field +2.5; shortstop +7.5; and left and right field −7.5. Pitcher position is zero. A designated hitter receives −17.5 per 162 exposure-derived game equivalents. Those equivalents use actual designated-hitter plate appearances divided by the relevant league-season mean designated-hitter plate appearances per team-game with a designated hitter. If that reference is absent, the declared ordinary-plate-appearance reference is used. Pinch hitters and pinch runners receive no position allocation without actual defensive or designated-hitter exposure. This fixed schedule is applied across history as a comparison convention.

Three league-season pools are centered separately. Ordinary batting, running, nonpitcher fielding and position use ordinary plate appearances as exposure. Pitcher-role batting and running use pitcher-role plate appearances. Pitching plus own pitcher-fielding uses pitching outs, separately for RA and FIP. If a pool has raw total $T$, player $i$ has exposure $x_i$, and total exposure is $X$, its correction is

$$\text{centering}_i = -T\frac{x_i}{X}. \quad \text{(S10)}$$

An empty pool requires no correction, and zero exposure receives no correction. The scored team-win target does not determine these adjustments.

The common league-season conversion is

$$\text{RPW} = 10\sqrt{\frac{R^{\text{adj}}/G_{\text{team}}}{4.5}}, \quad \text{(S11)}$$

where $R^{\text{adj}}$ is adjusted league scoring and $G_{\text{team}}$ counts team-games, including both sides of each game. The replacement budget is $0.206G_{\text{team}}$ wins: 57% is allocated by ordinary plate appearances and 43% by pitching outs. Two-way players may receive both allowances for distinct workloads. A zero-out pitcher receives no outs-based replacement credit while retaining the above-average contribution or debit from the appearance.

With complete finite inputs, centering makes each declared pool sum to zero. Therefore

$$\sum_{\text{teams}} \text{WAR} = 0.206\sum_{\text{teams}} G,$$
$$\sum_{\text{teams}} (0.294G + \text{WAR}) = 0.5\sum_{\text{teams}} G. \quad \text{(S12)}$$

A complete opposing-game inventory also assigns half of all team-games as credited wins. This explains why the pooled implied-win bias is near zero on a complete matched scope. It leaves individual team errors, MAE and RMSE undetermined. An incomplete comparison sample or equal-row per-162 weighting need not preserve zero bias.

S3. Provider scope and comparison samples

FanGraphs team WAR is all-position batting WAR, including pitcher batting, plus pitching-only WAR once. Baseball-Reference's supplied team Total is used once. Independently ranked role columns are joined by their own team code before role arithmetic is checked. Provider definitions differ from the iWAR conventions described above (Sports Reference, n.d.; Slowinski, 2012).

Source competition, complete-season records and required roles must reconcile before a team-season enters the comparison. Provider tables do not supply complete game-identity lists. The study certifies the iWAR game identities and checks provider roles and published records, but aggregate agreement alone cannot establish identical underlying game membership across providers. Baseball-Reference game and tie counts that come from the canonical game logs are identified as such; they are not described as independent provider observations.

**Table S1. Comparison samples.** Replay years refer to the target season. The missing-fWAR partition is classified by the origin season in replay.

| Sample | Same-season team-seasons | Same-season years | Consecutive franchise pairs | Replay target years |
|---|---|---|---|---|
| Common four-WAR sample | 1,471 | 79 | 1,441 | 78 |
| Broader bWAR sample | 2,554 | 118 | 2,524 | 117 |
| Broader sample: common partition A | 1,471 | 79 | 1,441 | 78 |
| Broader sample: missing-fWAR partition B | 1,068 | 39 | 1,068 | 39 |
| Broader sample: other eligible partition C | 15 | — | 15 | — |

The common same-season and origin years are 1908–1967, 1975, 1988–1997, 2017–2019 and 2021–2025. Full fWAR roles are unavailable in 1968–1974, 1976–1987, 1998–2016 and 2020. The expanded common sample adds 69 years to the original comparison: 59 early years excluding the original 1927 example, plus 1988–1997. It adds no origin year after 1997. Equal-season weighting changes the weight assigned to observed years; it cannot supply the missing periods.

The broader bWAR sample starts with all 2,568 complete conventional products. Fourteen fail the additional complete-record comparison because their event-derived and canonical credited wins differ. Their event and canonical game counts agree, and their canonical game identities are complete. Table S2 retains the conflicting records without assigning an unverified cause.

**Table S2. Complete-product exclusions at the record-reconciliation stage.** Credited wins equal $W + T / 2$. Team codes are the source's historical club identifiers.

| Year and club | Event credited wins | Canonical credited wins | Event ties | Canonical ties |
|---|---|---|---|---|
| 1909 NY1 | 94.0 | 94.5 | 6 | 5 |
| 1909 PHI | 75.0 | 74.5 | 2 | 1 |
| 1916 BRO | 94.5 | 95.0 | 3 | 2 |
| 1916 CHN | 69.0 | 68.5 | 3 | 2 |
| 1917 CHA | 100.5 | 101.0 | 3 | 2 |
| 1917 CLE | 89.5 | 89.0 | 3 | 2 |
| 1941 BOS | 83.5 | 84.5 | 1 | 1 |
| 1941 WS1 | 72.0 | 71.0 | 2 | 2 |
| 1942 BSN | 59.0 | 60.0 | 2 | 2 |
| 1942 NY1 | 87.0 | 86.0 | 2 | 2 |
| 1971 NYA | 81.0 | 82.0 | 0 | 0 |
| 1971 WS2 | 64.0 | 63.0 | 0 | 0 |
| 1974 CLE | 77.5 | 77.0 | 1 | 0 |
| 1974 TEX | 84.0 | 84.5 | 2 | 1 |

Thus 2,594 potential products minus 26 incomplete products minus 14 record exclusions gives 2,554 broader comparison rows. A separate set of 47 unresolved player-bulk versus supplied-team decompositions is excluded from player agreement. Those discrepancies do not remove a valid supplied team Total when its own team-level checks pass, and the 47 cases must not be treated as another disjoint subtraction from Table S1.

## S4. Targets, run comparators and calibration

### S4.1. Same-season reconstruction and franchise replay

For a team-season with $G = W + L + T$, the outcome is credited win fraction

$$y = \frac{W + T/2}{G}. \quad \text{(S13)}$$

The fixed WAR conversion is $p = 0.294 + \text{WAR}/G$. Same-season reconstruction compares this quantity with that season's $y$. Franchise replay uses origin-year WAR divided by origin-year games to predict the next year's credited win fraction. Only verified consecutive-year franchise links are eligible. Relocations and renamings require an explicit franchise mapping; equal team codes alone do not establish continuity. Missing years are not bridged. A target needs a complete record but does not need a target-year WAR value.

Primary errors are $e = 162(p - y)$, in credited wins per 162 played games. Actual-season companion errors multiply the fraction difference by the scored season's games, which are target-year games for replay. The target's games enter the outcome and reporting denominator, not the predictor. Predictions remain unclipped. Replay summarizes persistence in franchise-level performance with current-vintage retrospective inputs; player aging, trades and roster changes are not modeled.

S4.2. Run-based references

The run comparators use reconciled, unadjusted complete-season game-log runs scored $RS$ and runs allowed $RA$. This is a different input convention from the automatic-runner-adjusted valuation references. The Pythagorean exponent-two fraction is

$$p_2 = \frac{RS^2}{RS^2 + RA^2}. \quad \text{(S14)}$$

The fixed PythagenPat fraction uses

$$\gamma = \left(\frac{RS + RA}{G}\right)^{0.285},$$

$$p_{\text{PP}} = \frac{RS^{\gamma}}{RS^{\gamma} + RA^{\gamma}}. \quad \text{(S15)}$$

The exponent rule is fixed; it is not chosen using this study's outcomes. If one run total is zero and the other positive, the formula uses the corresponding limit. Both-zero, negative, nonfinite or nonpositive-game inputs are unavailable. A linear run-differential sensitivity uses $0.5 + (RS - RA)/(G \ \text{RPW})$ where the saved conversion and matching source scope support it. Its calibrated predictor is raw run differential per game. The constant .500 reference is always retained, and previous credited record is an additional replay reference.

Every comparison uses identical eligible rows for all included methods. If a run comparator requires an additional intersection, all WAR methods are rescored on that same intersection. Event-run sensitivities retain their distinct source labels.

S4.3. Scalar calibration

Calibration estimates an intercept and slope for one declared predictor:

$$y = \alpha + \beta x + \varepsilon. \quad \text{(S16)}$$

Training uses ordinary least squares with equal row weights. Predictors are WAR per game for each WAR method, the corresponding Pythagorean or PythagenPat fraction, raw run differential per game, and previous record for replay. All methods share the same training and scored rows for a given comparison. A fold without predictor variance is unavailable for every calibrated method on that fold, preserving the common scored rows. The .500 reference remains fixed. Same-season calibration leaves the entire scored calendar year out and trains on the other eligible years. Later years may therefore inform an earlier scored year. This measures the effect of changing the conversion from a summary to wins. Replay calibration trains only on strictly earlier completed target years and requires at least ten such years and 150 eligible pairs. Its initial years provide training and are absent from the calibrated evaluation.

**Table S3. Calibration populations.** Each fixed method is also rescored on the identical calibrated evaluation rows.

| Comparison | Training rule | Scored rows | Scored years |
| --- | --- | --- | --- |
| Common same-season | Leave entire year out | 1,471 | 79 |
| Common replay | Earlier target years only | 1,282 | 67 |
| Broader bWAR same-season | Leave entire year out | 2,554 | 118 |
| Broader bWAR replay | Earlier target years only | 2,359 | 106 |

The broader comparison receives coefficients fitted to its own eligible training rows. A previously fitted coefficient could be reused only if its task, method, ordered training identities, predictor values and target fractions were identical. All achieved broader training samples differed, so their coefficients were fitted separately. This distinction prevents a change in sample size from being confused with a change in the training population. There are no era-specific fitted maps, tuned exponents, blends, clipping rules or outcome-selected iWAR variants.

## S5. Error summaries, intervals and display precision

The main analysis weights every eligible team-season or replay pair equally. The equal-season sensitivity assigns each scored year equal total weight and divides that year's weight equally among its teams. MAE is weighted mean absolute error; RMSE is the square root of weighted mean squared error; bias is weighted mean signed error. RMSE is calculated from all row-level squared errors, not by averaging annual RMSEs.

Prediction $R^2 = 1 - \mathrm{SSE} / \mathrm{SST}$ uses the weighted outcome mean and the declared error units. Pearson correlation and its square are reported separately. Spearman correlation is computed from weighted tied ranks: each tied value receives the cumulative mass below it plus half the tied mass, followed by weighted Pearson correlation. With equal row weights this reduces to ordinary average tied ranks. Constant or insufficient samples return missing statistics. None of these quantities is a percentage of games correctly predicted.

Paired effects subtract the provider's error from the corresponding iWAR error. Negative differences favor iWAR. The four primary contrasts compare each iWAR version with each provider. The broader analysis compares each iWAR version with bWAR. All declared contrasts remain visible regardless of their sign.

S5.1. Calendar-block resampling

Intervals use 5,000 paired resamples of nonoverlapping calendar blocks. Blocks are anchored at 1908; five years is the primary length, with one-year and ten-year sensitivities. Replay blocks are assigned by target year. Calendar gaps remain gaps. If there are $B$ nonempty blocks, a draw selects exactly $B$ blocks with replacement and retains every paired row in each selected copy. Methods share the same sampled blocks, so their losses are compared on the same observations. For equal-season weighting, each sampled copy of a year receives its own equal total weight. MAE and RMSE are recomputed on each draw. The interval endpoints are the 2.5th and 97.5th percentiles using linear interpolation. The common analysis uses saved seed 23023 and the broader analysis uses seed 24024. Fewer than ten nonempty blocks results in a reported point estimate and block count with the interval withheld. All primary five-year-block era intervals are withheld under this rule; some one-year sensitivity intervals have sufficient support.

The intervals describe variation across resampled historical blocks with the observed inputs, fitted methods and saved coefficients held fixed. They do not include uncertainty from method development, retrospective source revision, fitted valuation inputs or coefficient estimation. Nonstationarity over baseball history and dependence across block boundaries remain possible. The intervals are pointwise across the stated contrasts and do not receive a simultaneous multiplicity adjustment. These conditions should accompany any interpretation of an interval that excludes zero. Block resampling is motivated by dependence among observations; it does not make historical periods exchangeable in all respects (Künsch, 1989).

Leave-one-year-out influence summaries hold predictions and coefficients fixed and remove one scored year at a time. The remaining analyses retain the declared five eras, AL/NL splits, original versus added years, and workload sensitivities requiring at least 140 games in the relevant season or both replay seasons. They are reported without selecting a favorable era or convention.

S5.2. Rounding bounds

Provider display precision is treated separately from statistical uncertainty. For a WAR display half-step bound $b_i$, the fixed-map bound on an error in wins per 162 is $162 b_i / G_i$, using origin games for replay. FanGraphs' combined total adds the batting and pitching half-step bounds. Baseball-Reference's supplied Total has its own bound. The conditional calibrated bound multiplies by the absolute saved slope. It does not include the effect of rounding training inputs on the fitted slope and intercept.

The weighted mean row bound bounds the change in MAE or bias. The square root of the weighted mean squared row bound bounds the change in RMSE. A paired-effect bound adds the component bounds. These are limits attributable to displayed precision under the stated calculation, and they are not model-error intervals. For example, the common fixed replay FIP–fWAR MAE difference is −0.043 wins per 162, smaller than its paired display-rounding bound of ±0.104 and accompanied by a conditional resampling interval spanning zero. The result does not establish that the two underlying methods have equal forecasting performance.

S6. Descriptive run-accounting comparison

The accounting comparison asks how each WAR-based prediction differs from the game-log PythagenPat prediction on the same rows. Let $e_{\mathrm{run}} = 162(p_{\mathrm{PP}} - y)$ and $d = 162(p_{\mathrm{WAR}} - p_{\mathrm{PP}})$. For normalized weights,

$$\mathrm{MSE}_{\mathrm{WAR}} - \mathrm{MSE}_{\mathrm{PP}} = E_w[d^2] + 2E_w[e_{\mathrm{run}} d]. \quad \text{(S17)}$$

The first term measures the squared distance between predictions. The second describes whether that difference tends to add to or offset the run-based prediction's error. This is an identity for squared error. It does not assign the difference to batting, pitching or fielding and does not supply an additive decomposition of MAE or RMSE.

**Table S4. Common-cohort same-season MSE accounting.** Units are squared credited wins per 162; values are rounded to three decimals.

| Method | Squared prediction distance | Error cross-term | Excess MSE over PythagenPat |
|---|---|---|---|
| iWAR RA | 7.050 | −2.387 | 4.664 |
| iWAR FIP | 11.568 | −3.804 | 7.764 |
| bWAR | 21.210 | −3.815 | 17.396 |
| fWAR | 22.046 | −3.741 | 18.305 |

The saved full-precision identity residual is at most $2.13 \times 10^{-14}$. The arithmetic is exact up to numerical accumulation, while its interpretation remains descriptive. A complete matched provider component ledger was unavailable. Consequently, the observed differences cannot be causally assigned to particular provider adjustments.

## S7. Agreement with player bWAR

### S7.1. Identity, roles and exposure

The player supplement compares iWAR with current-vintage Baseball-Reference WAR. Exact unique Retrosheet-to-Baseball-Reference identifiers come from the Chadwick Baseball Bureau Persons Register, commit `781c175e748b89e5730fd3f5de0086b9355b2f92`, dated October 2, 2026 (Chadwick Baseball Bureau, 2026). The public extract is updated periodically and can lag the full register. Empty, duplicated or ambiguous identifiers are excluded. Names, observed WAR and player prominence do not determine matches.

The primary unit is a player-team-season. Disjoint batting stints, including pitcher batting, are added once, and pitching-only stints are added once. A present missing WAR value remains missing. A truly absent role can contribute zero only when complete-file evidence and zero role exposure establish its absence. Pitching-only comparisons can use a finite pitching component independently of a missing batting component that is irrelevant to that comparison. All 47 unresolved bulk-player versus supplied-team decompositions are excluded from the player analysis.

Complete player-season aggregation requires every observed AL/NL club stint in the union of both sources to be eligible. An observed ineligible stint cannot disappear when the other stints are summed. Cross-league values are summed in wins after the original league-specific conversions. This yields 90,891 combined player-team rows and 83,642 complete union player-seasons across all 118 seasons.

The batting-rate comparison requires zero pitching outs in both sources and at least 100 plate appearances in each. Rates use each source's own plate appearances and are expressed per 600 PA. The pitching-rate comparison requires at least 90 literal outs in each source and uses $600 \times \mathrm{WAR} / \mathrm{outs}$, equivalent to WAR per 200 innings. These rates describe workload-normalized agreement; they do not project a player to a future workload.

## S7.2. Results

**Table S5. Player-team agreement with bWAR.** Mean difference is iWAR minus bWAR. Total values and MAE are in wins; rate rows use their stated rate units. Rank ρ is Spearman rank correlation. All rows represent 118 seasons.

| Comparison | Basis | N | Pearson r | Rank ρ | Mean difference | MAE |
|---|---|---|---|---|---|---|
| Combined total | RA | 90,891 | 0.934 | 0.882 | 0.003 | 0.406 |
| Combined total | FIP | 90,891 | 0.916 | 0.844 | 0.002 | 0.449 |
| Batting rate per 600 PA | RA | 35,095 | 0.895 | 0.887 | −0.035 | 0.854 |
| Batting rate per 600 PA | FIP | 35,095 | 0.895 | 0.887 | −0.035 | 0.854 |
| Pitching-only total | RA | 48,136 | 0.902 | 0.850 | 0.038 | 0.451 |
| Pitching-only total | FIP | 48,136 | 0.854 | 0.752 | 0.038 | 0.552 |
| Pitching rate per 200 IP | RA | 29,518 | 0.800 | 0.781 | 0.280 | 1.330 |
| Pitching rate per 200 IP | FIP | 29,518 | 0.674 | 0.660 | 0.256 | 1.619 |

**Table S6. Complete union player-season agreement with bWAR.** Eligibility is required across all observed club stints before aggregation. Units, correlation definitions and difference direction follow Table S5.

| Comparison | Basis | N | Pearson r | Rank ρ | Mean difference | MAE |
|---|---|---|---|---|---|---|
| Combined total | RA | 83,642 | 0.933 | 0.886 | 0.002 | 0.428 |
| Combined total | FIP | 83,642 | 0.916 | 0.851 | 0.001 | 0.472 |
| Batting rate per 600 PA | RA | 34,285 | 0.897 | 0.890 | −0.033 | 0.842 |
| Batting rate per 600 PA | FIP | 34,285 | 0.897 | 0.890 | −0.033 | 0.842 |
| Pitching-only total | RA | 33,045 | 0.900 | 0.859 | 0.048 | 0.506 |
| Pitching-only total | FIP | 33,045 | 0.850 | 0.769 | 0.045 | 0.620 |
| Pitching rate per 200 IP | RA | 23,103 | 0.806 | 0.785 | 0.310 | 1.287 |
| Pitching rate per 200 IP | FIP | 23,103 | 0.680 | 0.662 | 0.287 | 1.576 |

RA agrees more closely with bWAR than FIP in these comparisons. Total agreement includes shared variation in playing time. The separately defined pitching-rate cohorts have lower correlations, but they also have different eligibility and units, so the difference does not isolate a causal playing-time effect. Component disagreements can offset each other in a close total, and neither method supplies an observed label for individual contribution.

Mean provider-minus-iWAR exposure differences in the combined player-team sample are −0.000462 PA and −0.000011 literal outs. Small mean differences do not demonstrate exact event-category agreement. Source exposure values and discrepancies remain recorded, including differing historical PA conventions. No uniform player-file rounding unit was established, so this supplement does not impose a numerical player rounding bound. The all-era, original-versus-added-year and workload-stratified results remain in the aggregate player table. No representative free fWAR player-bulk sample was obtained for this analysis.

## S8. Additional tables and reproduction scope

The accompanying aggregate tables retain all declared analyses. Table S7 identifies their contents so that readers can find a result without tracing the project's development history. The machine-readable tables preserve full numerical precision; printed tables round results for readability.

**Table S7. Machine-readable supplementary table index.** Paths refer to the accompanying evidence directory.

| File | Contents |
| --- | --- |
| tables/common_metrics.csv | Four-method common-sample errors, correlations, prediction $R^2$, counts and prediction-range checks; fixed and calibrated conversions, units, weights and strata |
| tables/common_paired_effects.csv | Common-sample iWAR-minus-provider effects and conditional calendar-block intervals |
| tables/common_sensitivity_effects.csv | Declared common-sample sensitivity effects |

| File | Contents |
|---|---|
| tables/broader_metrics.csv | Broader bWAR-sample metrics and sample partitions under the declared conversions and weights |
| tables/broader_paired_effects.csv | Broader iWAR-minus-bWAR effects and conditional intervals |
| tables/broader_year_influence.csv | Removal of each scored year with predictions and coefficients held fixed |
| tables/common_rounding.csv and tables/broader_rounding_summary.csv | Provider display-precision bounds under the declared conversions |
| tables/coverage_matrix.csv and tables/coverage_absence_reasons.csv | Source and feature availability, comparison coverage and absence reasons |
| tables/run_accounting_bridge.csv | Full-precision terms in Equation S17 by declared scope and weight |
| tables/original_current_added_bridge.csv and tables/original_bridge_contributions.csv | Original and newly added comparison groups and their aggregate contributions |
| tables/player_aggregate_agreement.csv | Player agreement by role, workload, era, original/added years and aggregation unit |
| tables/player_aligned_role_bridge.csv | Saved aggregate role-scope comparisons |
| tables/common_out_of_range.csv | Unclipped prediction-range checks |

The expanded evidence package contains the saved aggregate results, protocols, conventional specification and the genuine executable scientific implementation with its recorded environment declarations. Figures are supplied separately. Offline aggregate reporting is reproducible without private inputs. Repeating the valuation and benchmark from source records additionally requires compatible authorized Retrosheet/provider observations, retained row membership and the documented dependencies. Provider-valued rows, exact predictions and raw downloads are excluded from distributable assets; missing observations remain missing.

The companion explorer and historical series simulations use separate targets and assumptions. They provide a way to inspect teams and conditional matchups but add no evidence to the team-WAR losses in this article. Experimental fielding studies are also outside the present comparison: their numerical results are not used to support the conventional estimator's individual-attribution claims.

## S9. Amended chronological learned-model study

### S9.1. Scope and information timing

This section reports the completed neural_team_wins_v01 experiment and its saved implementation amendment. It does not reopen model development. Reconstruction (R) predicts the completed season's credited winning fraction, (W + 0.5T)/(W + L + T), from that season's statistics. Franchise replay (F) uses season t−1 to predict season t through verified franchise continuity; expansion or missing/nonconsecutive links are excluded. Target exposure determines the label and actual-season scoring only; it is absent from replay features. Team seasons with unresolved credited records remain excluded. Prepared source coverage was 1908–2025: 2,554 reconstruction rows and 2,524 replay pairs before the post-1969 evaluation cut. The amended primary evaluation retains 1,558 R rows and 1,552 F pairs across 56 target years, 1970–2025. Current-vintage historical observations are not archived preseason inputs, and some historical outcomes had informed earlier project work.

### S9.2. Feature dictionary and preprocessing

CORE has 21 reconstruction features and 22 replay features. Rates use origin-season denominators in replay. Its groups are runs scored/game; runs allowed/game; batting hits, home runs, walks and strikeouts per plate appearance; pitching hits, home runs, walks and strikeouts allowed per batter faced; plate appearances/game; batters faced/game; log(1 + games); AL indicator; source league run environment; and the six optional event rates below. Replay adds prior credited winning fraction. The optional gate was determined from pre-1970 development completeness only: doubles/PA, triples/PA, HBP/PA, stolen bases/PA, caught stealing/PA and pitching HBP/BF all passed the 95% rule at 100% development coverage. AUGMENTED appends six saved conventional component rates before role-pool centering: batting, running, fielding, position, RA pitching and FIP pitching, each divided by source games. Separate saved centering terms are excluded, and position uses the saved raw position component. It does not include total WAR or replacement credit. RA and FIP pitching are separate features in this conditional team-label exercise; they are not added to define a new WAR total.

For every development fold and annual training set, the saved transformer fits feature medians, missingness indicators and standardization using training observations only. Missingness indicators are supplied for every feature. The corrected numerical-constant rule described below sets a constant feature to zero in every transformed observation, including future values. Medians, scales and constant masks never use evaluation rows. CORE already includes observed scoring and prevention; component tests concern additional conditional information, not whether component accounting is useful in general.

### S9.3. Selection, annual fitting and model settings

The three development folds trained through 1939, 1949 and 1959 and validated on the following decade through 1969. Selection pooled validation squared error. The neural development ensemble used seeds 240101 and 240102. Ties within $10^{-8}$ selected fewer parameters, then stronger regularization, then listed order. Each family, target and feature set received separate selection. The primary non-neural comparator, Ridge AUGMENTED, was designated from pre-1970 replay development and was retained after observing evaluation rankings.

The candidate multilayer perceptrons used hidden sizes 32/16 or 64/32, each crossed with the two weight decays. Each has two ReLU hidden layers, dropout 0.1 after the first, and a linear output predicting winning fraction minus 0.5. AdamW used learning rate 0.001, weight decay 0.001 or 0.01, full-batch float32 training, a maximum of 400 epochs, a minimum of 50 and patience 30 with improvement threshold $10^{-6}$. The actual best-validation weights were retained.

Annual epochs were the rounded median of the six development best epochs, clipped to 50–400. Each annual ensemble averaged seeds 240201, 240202 and 240203; annual refits did not early-stop on evaluation outcomes. Ridge tested penalties 0.1, 1, 10 and 100. Histogram gradient boosting tested 7 or 15 leaves and L2 penalties 1 or 10, with learning rate 0.05, 200 iterations, minimum leaf size 20, no early stopping and seed 240301. Fixed published configuration files preserve all candidates and tie rules.

**Table S8. Corrected selected settings.** Epochs apply only to the neural model. Settings for CORE were preserved from the original run after reuse checks; AUGMENTED was reselected on the same development grid after the implementation repair.

| Target | Features | Family | Selected configuration | Epochs |
|---|---|---|---|---|
| F | CORE | MLP | 64/32; decay 0.01 | 64 |
| F | CORE | Ridge | alpha 100 | — |
| F | CORE | HGB | 7 leaves; L2 10 | — |
| F | AUGMENTED | MLP | 64/32; decay 0.01 | 50 |
| F | AUGMENTED | Ridge | alpha 100 | — |
| F | AUGMENTED | HGB | 7 leaves; L2 10 | — |
| R | CORE | MLP | 32/16; decay 0.001 | 384 |
| R | CORE | Ridge | alpha 1 | — |
| R | CORE | HGB | 7 leaves; L2 10 | — |
| R | AUGMENTED | MLP | 32/16; decay 0.01 | 253 |
| R | AUGMENTED | Ridge | alpha 1 | — |
| R | AUGMENTED | HGB | 7 leaves; L2 10 | — |

Every annual R and F fit used only earlier completed target years. R input statistics still describe the scored season, so chronological model fitting does not make reconstruction a pregame forecast. This annual R policy differs from the original manuscript's leave-one-year-out scalar reconstruction calibration. The learned families used full eligible internal earlier history. Saved scalar maps required at least ten earlier target years and 150 pairs and retained their own cohort-specific histories. Fixed formulas and the .500 reference were not refitted. An evaluation-row intersection does not equalize these training histories. Separately selected CORE and AUGMENTED settings make their contrast an effect under a selection procedure, rather than a fixed-setting ablation.

S9.4. Numerical correction and chronology

The original detector in models.py used exact equality, scale == 0, to decide whether a training feature was constant. Conventional position weights sum to zero in the pre-designated-hitter team accounting when defensive exposures agree. Round-off nevertheless left position/game values of approximately $-2.36\times10^{-17}$ to $+1.95\times10^{-17}$ in the 1,066-row training set through source year 1972, with standard deviation approximately $7.22\times10^{-18}$. A genuine DH-era position rate near −0.11 then became a standardized input near $-1.6\times10^{16}$. The first affected annual comparisons were R1973 and F1974, particularly the twelve AL teams. The saved diagnosis found a numerical constant-detection defect, rather than a change in source convention between the original and expansion seasons. Initial catastrophic errors were invalid software output, not evidence about model quality.

The new repair_constants.py applies one generic training-only rule: a feature is numerically constant when its range is at most 16 × float64 machine epsilon × max(1, |minimum|, |maximum|). Float64 epsilon is $2.220446049250313\times10^{-16}$. Its scale is set to one and all transformed values of that feature are zero. The guard follows machine precision and the mathematical zero, independently of which model's error improved; it preserves meaningful small variation. It does not alter the component definition, use evaluation-derived thresholds, clip predictions, or add a year/feature exception. The original exact-zero code remains available alongside the corrected implementation.

Initial post-1969 evaluation was inspected before diagnosis. The retained constant_diagnosis.json, NUMERICAL_REPAIR.md and pre_repair_01 archive document that sequence; the available record does not establish an exact wall-clock timestamp for first human inspection, so none is inferred. The implementation amendment was frozen at 2026-10-18T00:55:15.428446 UTC, after that inspection and before corrected annual fitting. Corrected AUGMENTED development reused the same grid, folds, seeds and tie rules. The R AUGMENTED neural epoch selection changed from 268 to 253; the corrected selected settings appear in Table S8. The repair caused 96 AUGMENTED development invocations, 560 annual invocations and one diagnostic transformer fixture invocation. Those are part of the completed study, not new publication-stage work.

Reuse checks established exact equality of CORE transformed matrices for the development/annual cases and retained 184,522 CORE/scalar/fixed prediction rows. The repair replaced 87,096 AUGMENTED rows within the 271,618-row long prediction store, with labels and metadata preserved. Original outputs and receipts remain in pre_repair_01, with the before-repair evidence preserved; corrected summaries, selection, predictions and reports occupy separate current/repair_01 files and have separate fingerprints. Main Table 5 and the new tables below use corrected predictions. The pre-existing conventional historical benchmark and its tables were not changed by this repair. Corrected development freezing does not remove knowledge gained from initial evaluation: the study is explicitly amended historical evidence.

### S9.5. Cohorts, metrics and interval applicability

Tables S9–S20 summarize fixed saved predictions assembled for publication. Stable cohort identifiers and membership fingerprints appear in cohort_inventory.csv; private row-key lists

remain in the retained local evidence/private directory and are excluded from distributable archives. Original primary membership is retained separately. A_internal is the original post-1969 internal evaluation. B_four_WAR intersects actual iWAR RA, iWAR FIP, bWAR, fWAR and learned-model outputs. C_broader_bWAR omits the fWAR requirement; its post-1969 rows happen to equal internal membership, but its provider calibration retains its own training policy. Calibrated suffixes require all displayed saved scalar maps to be present, alongside unchanged learned predictions and the fixed .500 reference. These publication-stage tables do not replace the original primary comparison.

Each table uses equal team-season weights, unbounded predictions and credited wins per 162. Bias is predicted minus observed wins. MAE is weighted absolute error; RMSE is the square root of pooled weighted squared error, not the average of annual RMSEs. Prediction $R^2$ is 1 − SSE/SST about the weighted observed mean; it is separate from Pearson r squared. Pearson r and Spearman rho are correlations, not classification accuracy. Undefined values remain explicitly unavailable. All methods in a displayed table have the same rows, outcomes, units and weights. F features use source-season games; target games appear only in labels and actual-season sensitivities. Ties receive half a credited win.

Saved intervals are used only when target, method contrast, rows, weighting, units and sensitivity match exactly. The neural protocol used 2,000 paired draws of nonoverlapping five-calendar-year blocks anchored at 1970, with the final 2025 endpoint forming a one-year block, seed 240401. Fewer than ten nonempty blocks causes interval withholding. Predictions were held fixed, giving conditional pointwise intervals that omit selection, refitting and prior-development uncertainty. Provider intersections and newly summarized Ridge/HGB component differences have no borrowed intervals. Saved summaries and draw indices remain available locally; no new indices or bootstrap draws were generated for this package.

**Table S9. Next-season replay: internal.** N = 1,552; 56 target years, 1970–2025. Scalar entries use fixed formulas; learned entries use annual expanding fits. All errors are credited wins per 162. Detailed machine files also identify feature set, mapping, weighting, training policy, interval applicability and membership hash.

| Method | MAE | RMSE | Bias | Pred. $R^2$ | r | rho |
|---|---|---|---|---|---|---|
| .500 | 9.519 | 11.613 | −0.089 | −0.000 | unavailable | unavailable |
| HGB AUG | 8.231 | 10.138 | 0.571 | 0.238 | 0.501 | 0.487 |
| HGB CORE | 8.178 | 10.119 | 0.639 | 0.241 | 0.504 | 0.494 |
| MLP AUG | 8.176 | 10.163 | −0.010 | 0.234 | 0.497 | 0.495 |
| MLP CORE | 8.095 | 9.999 | 0.685 | 0.259 | 0.514 | 0.503 |
| Previous record | 9.257 | 11.564 | −0.079 | 0.008 | 0.508 | 0.493 |
| PythagenPat | 8.856 | 11.112 | −0.043 | 0.084 | 0.525 | 0.507 |
| Pythagorean (2) | 9.159 | 11.496 | −0.025 | 0.020 | 0.525 | 0.507 |
| Ridge AUG | 7.989 | 9.946 | 0.457 | 0.266 | 0.533 | 0.518 |
| Ridge CORE | 7.989 | 9.935 | 0.445 | 0.268 | 0.534 | 0.519 |
| iWAR FIP | 8.684 | 10.887 | −0.084 | 0.121 | 0.541 | 0.522 |
| iWAR RA | 9.098 | 11.418 | −0.077 | 0.033 | 0.515 | 0.496 |

**Table S10. Next-season replay: internal_calibrated.** N = 1,552; 56 target years, 1970–2025. Scalar entries use their saved annual expanding calibrations; learned entries are unchanged; .500 remains fixed. All errors are credited wins per 162. Detailed machine files also identify feature set, mapping, weighting, training policy, interval applicability and membership hash.

| Method | MAE | RMSE | Bias | Pred. $R^2$ | r | rho |
|---|---|---|---|---|---|---|
| .500 | 9.519 | 11.613 | −0.089 | −0.000 | unavailable | unavailable |
| HGB AUG | 8.231 | 10.138 | 0.571 | 0.238 | 0.501 | 0.487 |
| HGB CORE | 8.178 | 10.119 | 0.639 | 0.241 | 0.504 | 0.494 |
| MLP AUG | 8.176 | 10.163 | −0.010 | 0.234 | 0.497 | 0.495 |

| Method | MAE | RMSE | Bias | Pred. $R^2$ | r | rho |
|---|---|---|---|---|---|---|
| MLP CORE | 8.095 | 9.999 | 0.685 | 0.259 | 0.514 | 0.503 |
| Previous record | 8.168 | 10.090 | 0.169 | 0.245 | 0.507 | 0.492 |
| PythagenPat | 8.041 | 9.983 | 0.184 | 0.261 | 0.525 | 0.507 |
| Pythagorean (2) | 8.044 | 9.983 | 0.188 | 0.261 | 0.524 | 0.507 |
| Ridge AUG | 7.989 | 9.946 | 0.457 | 0.266 | 0.533 | 0.518 |
| Ridge CORE | 7.989 | 9.935 | 0.445 | 0.268 | 0.534 | 0.519 |
| Run differential | 8.039 | 9.986 | 0.174 | 0.260 | 0.524 | 0.508 |
| iWAR FIP | 7.956 | 9.906 | 0.166 | 0.272 | 0.540 | 0.522 |
| iWAR RA | 8.096 | 10.073 | 0.170 | 0.248 | 0.515 | 0.497 |

**Table S11. Next-season replay: four_WAR.** N = 504; 18 target years, 1976–2025. Scalar entries use fixed formulas; learned entries use annual expanding fits. All errors are credited wins per 162. Detailed machine files also identify feature set, mapping, weighting, training policy, interval applicability and membership hash.

| Method | MAE | RMSE | Bias | Pred. $R^2$ | r | rho |
|---|---|---|---|---|---|---|
| .500 | 9.637 | 12.015 | −0.128 | −0.000 | unavailable | unavailable |
| HGB AUG | 8.495 | 10.544 | 0.617 | 0.230 | 0.491 | 0.479 |
| HGB CORE | 8.548 | 10.597 | 0.789 | 0.222 | 0.486 | 0.474 |
| MLP AUG | 8.304 | 10.297 | 0.516 | 0.265 | 0.520 | 0.501 |
| MLP CORE | 8.308 | 10.272 | 0.711 | 0.269 | 0.523 | 0.496 |
| Previous record | 9.807 | 12.193 | −0.129 | −0.030 | 0.480 | 0.452 |
| PythagenPat | 9.197 | 11.527 | −0.062 | 0.080 | 0.528 | 0.497 |
| Pythagorean (2) | 9.480 | 11.908 | −0.031 | 0.018 | 0.528 | 0.497 |
| Ridge AUG | 8.247 | 10.316 | 0.458 | 0.263 | 0.531 | 0.506 |
| Ridge CORE | 8.265 | 10.321 | 0.437 | 0.262 | 0.530 | 0.506 |

| Method | MAE | RMSE | Bias | Pred. $R^2$ | r | rho |
|---|---|---|---|---|---|---|
| bWAR | 9.525 | 11.919 | −0.170 | 0.016 | 0.480 | 0.456 |
| fWAR | 8.915 | 11.291 | −0.164 | 0.117 | 0.534 | 0.509 |
| iWAR FIP | 8.917 | 11.287 | −0.129 | 0.118 | 0.543 | 0.511 |
| iWAR RA | 9.398 | 11.857 | −0.129 | 0.026 | 0.516 | 0.482 |

**Table S12. Next-season replay: four_WAR_calibrated.** N = 504; 18 target years, 1976–2025. Scalar entries use their saved annual expanding calibrations; learned entries are unchanged; .500 remains fixed. All errors are credited wins per 162. Detailed machine files also identify feature set, mapping, weighting, training policy, interval applicability and membership hash.

| Method | MAE | RMSE | Bias | Pred. $R^2$ | r | rho |
|---|---|---|---|---|---|---|
| .500 | 9.637 | 12.015 | −0.128 | −0.000 | unavailable | unavailable |
| HGB AUG | 8.495 | 10.544 | 0.617 | 0.230 | 0.491 | 0.479 |
| HGB CORE | 8.548 | 10.597 | 0.789 | 0.222 | 0.486 | 0.474 |
| MLP AUG | 8.304 | 10.297 | 0.516 | 0.265 | 0.520 | 0.501 |
| MLP CORE | 8.308 | 10.272 | 0.711 | 0.269 | 0.523 | 0.496 |
| Previous record | 8.671 | 10.712 | 0.053 | 0.205 | 0.478 | 0.450 |
| PythagenPat | 8.367 | 10.374 | 0.090 | 0.254 | 0.527 | 0.496 |
| Pythagorean (2) | 8.360 | 10.366 | 0.098 | 0.256 | 0.526 | 0.496 |
| Ridge AUG | 8.247 | 10.316 | 0.458 | 0.263 | 0.531 | 0.506 |
| Ridge CORE | 8.265 | 10.321 | 0.437 | 0.262 | 0.530 | 0.506 |
| Run differential | 8.373 | 10.390 | 0.067 | 0.252 | 0.526 | 0.495 |
| bWAR | 8.715 | 10.829 | 0.219 | 0.188 | 0.476 | 0.455 |
| fWAR | 8.338 | 10.516 | 0.124 | 0.234 | 0.532 | 0.508 |
| iWAR FIP | 8.268 | 10.330 | 0.059 | 0.261 | 0.542 | 0.511 |
| iWAR RA | 8.409 | 10.500 | 0.059 | 0.236 | 0.514 | 0.483 |

**Table S13. Next-season replay: broader_bWAR.** N = 1,552; 56 target years, 1970–2025. Scalar entries use fixed formulas; learned entries use annual expanding fits. All errors are credited wins per 162. Detailed machine files also identify feature set, mapping, weighting, training policy, interval applicability and membership hash.

| Method | MAE | RMSE | Bias | Pred. $R^2$ | r | rho |
|---|---|---|---|---|---|---|
| .500 | 9.519 | 11.613 | −0.089 | −0.000 | unavailable | unavailable |
| HGB AUG | 8.231 | 10.138 | 0.571 | 0.238 | 0.501 | 0.487 |
| HGB CORE | 8.178 | 10.119 | 0.639 | 0.241 | 0.504 | 0.494 |
| MLP AUG | 8.176 | 10.163 | −0.010 | 0.234 | 0.497 | 0.495 |
| MLP CORE | 8.095 | 9.999 | 0.685 | 0.259 | 0.514 | 0.503 |
| Previous record | 9.257 | 11.564 | −0.079 | 0.008 | 0.508 | 0.493 |
| PythagenPat | 8.856 | 11.112 | −0.043 | 0.084 | 0.525 | 0.507 |
| Pythagorean (2) | 9.159 | 11.496 | −0.025 | 0.020 | 0.525 | 0.507 |
| Ridge AUG | 7.989 | 9.946 | 0.457 | 0.266 | 0.533 | 0.518 |
| Ridge CORE | 7.989 | 9.935 | 0.445 | 0.268 | 0.534 | 0.519 |
| bWAR | 9.189 | 11.430 | −0.136 | 0.031 | 0.490 | 0.478 |
| iWAR FIP | 8.684 | 10.887 | −0.084 | 0.121 | 0.541 | 0.522 |
| iWAR RA | 9.098 | 11.418 | −0.077 | 0.033 | 0.515 | 0.496 |

**Table S14. Next-season replay: broader_bWAR_calibrated.** N = 1,552; 56 target years, 1970–2025. Scalar entries use their saved annual expanding calibrations; learned entries are unchanged; .500 remains fixed. All errors are credited wins per 162. Detailed machine files also identify feature set, mapping, weighting, training policy, interval applicability and membership hash.

| Method | MAE | RMSE | Bias | Pred. $R^2$ | r | rho |
|---|---|---|---|---|---|---|
| .500 | 9.519 | 11.613 | −0.089 | −0.000 | unavailable | unavailable |
| HGB AUG | 8.231 | 10.138 | 0.571 | 0.238 | 0.501 | 0.487 |

| Method | MAE | RMSE | Bias | Pred. $R^2$ | r | rho |
|---|---|---|---|---|---|---|
| HGB CORE | 8.178 | 10.119 | 0.639 | 0.241 | 0.504 | 0.494 |
| MLP AUG | 8.176 | 10.163 | −0.010 | 0.234 | 0.497 | 0.495 |
| MLP CORE | 8.095 | 9.999 | 0.685 | 0.259 | 0.514 | 0.503 |
| Previous record | 8.168 | 10.090 | 0.169 | 0.245 | 0.507 | 0.492 |
| PythagenPat | 8.041 | 9.983 | 0.184 | 0.261 | 0.525 | 0.507 |
| Pythagorean (2) | 8.044 | 9.983 | 0.188 | 0.261 | 0.524 | 0.507 |
| Ridge AUG | 7.989 | 9.946 | 0.457 | 0.266 | 0.533 | 0.518 |
| Ridge CORE | 7.989 | 9.935 | 0.445 | 0.268 | 0.534 | 0.519 |
| Run differential | 8.039 | 9.986 | 0.174 | 0.260 | 0.524 | 0.508 |
| bWAR | 8.312 | 10.283 | 0.286 | 0.216 | 0.489 | 0.478 |
| iWAR FIP | 7.956 | 9.906 | 0.166 | 0.272 | 0.540 | 0.522 |
| iWAR RA | 8.096 | 10.073 | 0.170 | 0.248 | 0.515 | 0.497 |

**Table S15. Same-season reconstruction: internal.** N = 1,558; 56 target years, 1970–2025. Scalar entries use fixed formulas; learned entries use annual expanding fits. All errors are credited wins per 162. Detailed machine files also identify feature set, mapping, weighting, training policy, interval applicability and membership hash.

| Method | MAE | RMSE | Bias | Pred. $R^2$ | r | rho |
|---|---|---|---|---|---|---|
| .500 | 9.554 | 11.656 | −0.010 | −0.000 | unavailable | unavailable |
| HGB AUG | 3.699 | 4.639 | 0.170 | 0.842 | 0.918 | 0.918 |
| HGB CORE | 3.408 | 4.323 | 0.069 | 0.862 | 0.929 | 0.930 |
| MLP AUG | 3.392 | 4.282 | 0.103 | 0.865 | 0.930 | 0.928 |
| MLP CORE | 3.340 | 4.253 | 0.213 | 0.867 | 0.932 | 0.929 |
| PythagenPat | 3.263 | 4.094 | 0.041 | 0.877 | 0.937 | 0.935 |
| Pythagorean (2) | 3.345 | 4.223 | 0.061 | 0.869 | 0.936 | 0.935 |

| Method | MAE | RMSE | Bias | Pred. $R^2$ | r | rho |
|---|---|---|---|---|---|---|
| Ridge AUG | 3.045 | 3.848 | 0.130 | 0.891 | 0.944 | 0.944 |
| Ridge CORE | 3.103 | 3.938 | −0.002 | 0.886 | 0.941 | 0.941 |
| iWAR FIP | 3.749 | 4.761 | −0.005 | 0.833 | 0.914 | 0.911 |
| iWAR RA | 3.760 | 4.733 | 0.002 | 0.835 | 0.917 | 0.916 |

**Table S16. Same-season reconstruction: internal_calibrated.** N = 1,558; 56 target years, 1970–2025. Scalar entries use their saved annual expanding calibrations; learned entries are unchanged; .500 remains fixed. All errors are credited wins per 162. Detailed machine files also identify feature set, mapping, weighting, training policy, interval applicability and membership hash.

| Method | MAE | RMSE | Bias | Pred. $R^2$ | r | rho |
|---|---|---|---|---|---|---|
| .500 | 9.554 | 11.656 | −0.010 | −0.000 | unavailable | unavailable |
| HGB AUG | 3.699 | 4.639 | 0.170 | 0.842 | 0.918 | 0.918 |
| HGB CORE | 3.408 | 4.323 | 0.069 | 0.862 | 0.929 | 0.930 |
| MLP AUG | 3.392 | 4.282 | 0.103 | 0.865 | 0.930 | 0.928 |
| MLP CORE | 3.340 | 4.253 | 0.213 | 0.867 | 0.932 | 0.929 |
| PythagenPat | 3.262 | 4.092 | 0.017 | 0.877 | 0.937 | 0.935 |
| Pythagorean (2) | 3.268 | 4.104 | 0.026 | 0.876 | 0.936 | 0.935 |
| Ridge AUG | 3.045 | 3.848 | 0.130 | 0.891 | 0.944 | 0.944 |
| Ridge CORE | 3.103 | 3.938 | −0.002 | 0.886 | 0.941 | 0.941 |
| Run differential | 3.319 | 4.152 | −0.004 | 0.873 | 0.935 | 0.934 |
| iWAR FIP | 3.802 | 4.823 | −0.017 | 0.829 | 0.914 | 0.911 |
| iWAR RA | 3.728 | 4.674 | −0.010 | 0.839 | 0.917 | 0.916 |

**Table S17. Same-season reconstruction: four_WAR.** N = 534; 19 target years, 1975–2025. Scalar entries use fixed formulas; learned entries use annual expanding fits. All errors are credited wins per 162. Detailed machine files also identify feature set, mapping, weighting, training policy, interval applicability and membership hash.

| Method | MAE | RMSE | Bias | Pred. $R^2$ | r | rho |
|---|---|---|---|---|---|---|
| .500 | 9.527 | 11.883 | 0.001 | −0.000 | unavailable | unavailable |
| HGB AUG | 3.617 | 4.563 | 0.193 | 0.853 | 0.924 | 0.924 |
| HGB CORE | 3.423 | 4.397 | 0.050 | 0.863 | 0.930 | 0.931 |
| MLP AUG | 3.288 | 4.211 | 0.262 | 0.874 | 0.935 | 0.932 |
| MLP CORE | 3.210 | 4.099 | 0.195 | 0.881 | 0.939 | 0.937 |
| PythagenPat | 3.181 | 4.041 | 0.078 | 0.884 | 0.942 | 0.939 |
| Pythagorean (2) | 3.309 | 4.222 | 0.112 | 0.874 | 0.941 | 0.939 |
| Ridge AUG | 3.029 | 3.862 | 0.192 | 0.894 | 0.947 | 0.946 |
| Ridge CORE | 3.110 | 3.962 | 0.043 | 0.889 | 0.945 | 0.943 |
| bWAR | 4.146 | 5.099 | −0.041 | 0.816 | 0.905 | 0.902 |
| fWAR | 4.249 | 5.380 | −0.035 | 0.795 | 0.894 | 0.887 |
| iWAR FIP | 3.654 | 4.641 | 0.000 | 0.847 | 0.922 | 0.919 |
| iWAR RA | 3.630 | 4.592 | −0.000 | 0.851 | 0.927 | 0.926 |

**Table S18. Same-season reconstruction: four_WAR_calibrated.** N = 534; 19 target years, 1975–2025. Scalar entries use their saved annual expanding calibrations; learned entries are unchanged; .500 remains fixed. All errors are credited wins per 162. Detailed machine files also identify feature set, mapping, weighting, training policy, interval applicability and membership hash.

| Method | MAE | RMSE | Bias | Pred. $R^2$ | r | rho |
|---|---|---|---|---|---|---|
| .500 | 9.527 | 11.883 | 0.001 | −0.000 | unavailable | unavailable |
| HGB AUG | 3.617 | 4.563 | 0.193 | 0.853 | 0.924 | 0.924 |
| HGB CORE | 3.423 | 4.397 | 0.050 | 0.863 | 0.930 | 0.931 |

| Method | MAE | RMSE | Bias | Pred. $R^2$ | r | rho |
|---|---|---|---|---|---|---|
| MLP AUG | 3.288 | 4.211 | 0.262 | 0.874 | 0.935 | 0.932 |
| MLP CORE | 3.210 | 4.099 | 0.195 | 0.881 | 0.939 | 0.937 |
| PythagenPat | 3.176 | 4.035 | 0.048 | 0.885 | 0.942 | 0.939 |
| Pythagorean (2) | 3.173 | 4.034 | 0.063 | 0.885 | 0.941 | 0.939 |
| Ridge AUG | 3.029 | 3.862 | 0.192 | 0.894 | 0.947 | 0.946 |
| Ridge CORE | 3.110 | 3.962 | 0.043 | 0.889 | 0.945 | 0.943 |
| Run differential | 3.256 | 4.108 | 0.005 | 0.880 | 0.940 | 0.938 |
| bWAR | 4.252 | 5.245 | 0.234 | 0.805 | 0.904 | 0.901 |
| fWAR | 4.541 | 5.752 | 0.081 | 0.766 | 0.893 | 0.887 |
| iWAR FIP | 3.779 | 4.783 | −0.008 | 0.838 | 0.923 | 0.919 |
| iWAR RA | 3.592 | 4.519 | −0.005 | 0.855 | 0.927 | 0.926 |

**Table S19. Same-season reconstruction: broader_bWAR.** N = 1,558; 56 target years, 1970–2025. Scalar entries use fixed formulas; learned entries use annual expanding fits. All errors are credited wins per 162. Detailed machine files also identify feature set, mapping, weighting, training policy, interval applicability and membership hash.

| Method | MAE | RMSE | Bias | Pred. $R^2$ | r | rho |
|---|---|---|---|---|---|---|
| .500 | 9.554 | 11.656 | −0.010 | −0.000 | unavailable | unavailable |
| HGB AUG | 3.699 | 4.639 | 0.170 | 0.842 | 0.918 | 0.918 |
| HGB CORE | 3.408 | 4.323 | 0.069 | 0.862 | 0.929 | 0.930 |
| MLP AUG | 3.392 | 4.282 | 0.103 | 0.865 | 0.930 | 0.928 |
| MLP CORE | 3.340 | 4.253 | 0.213 | 0.867 | 0.932 | 0.929 |
| PythagenPat | 3.263 | 4.094 | 0.041 | 0.877 | 0.937 | 0.935 |
| Pythagorean (2) | 3.345 | 4.223 | 0.061 | 0.869 | 0.936 | 0.935 |
| Ridge AUG | 3.045 | 3.848 | 0.130 | 0.891 | 0.944 | 0.944 |

| Method | MAE | RMSE | Bias | Pred. $R^2$ | r | rho |
|---|---|---|---|---|---|---|
| Ridge CORE | 3.103 | 3.938 | −0.002 | 0.886 | 0.941 | 0.941 |
| bWAR | 4.104 | 5.153 | −0.057 | 0.805 | 0.898 | 0.896 |
| iWAR FIP | 3.749 | 4.761 | −0.005 | 0.833 | 0.914 | 0.911 |
| iWAR RA | 3.760 | 4.733 | 0.002 | 0.835 | 0.917 | 0.916 |

**Table S20. Same-season reconstruction: broader_bWAR_calibrated.** N = 1,558; 56 target years, 1970–2025. Scalar entries use their saved annual expanding calibrations; learned entries are unchanged; .500 remains fixed. All errors are credited wins per 162. Detailed machine files also identify feature set, mapping, weighting, training policy, interval applicability and membership hash.

| Method | MAE | RMSE | Bias | Pred. $R^2$ | r | rho |
|---|---|---|---|---|---|---|
| .500 | 9.554 | 11.656 | −0.010 | −0.000 | unavailable | unavailable |
| HGB AUG | 3.699 | 4.639 | 0.170 | 0.842 | 0.918 | 0.918 |
| HGB CORE | 3.408 | 4.323 | 0.069 | 0.862 | 0.929 | 0.930 |
| MLP AUG | 3.392 | 4.282 | 0.103 | 0.865 | 0.930 | 0.928 |
| MLP CORE | 3.340 | 4.253 | 0.213 | 0.867 | 0.932 | 0.929 |
| PythagenPat | 3.262 | 4.092 | 0.017 | 0.877 | 0.937 | 0.935 |
| Pythagorean (2) | 3.268 | 4.104 | 0.026 | 0.876 | 0.936 | 0.935 |
| Ridge AUG | 3.045 | 3.848 | 0.130 | 0.891 | 0.944 | 0.944 |
| Ridge CORE | 3.103 | 3.938 | −0.002 | 0.886 | 0.941 | 0.941 |
| Run differential | 3.319 | 4.152 | −0.004 | 0.873 | 0.935 | 0.934 |
| bWAR | 4.155 | 5.226 | 0.167 | 0.799 | 0.898 | 0.896 |
| iWAR FIP | 3.802 | 4.823 | −0.017 | 0.829 | 0.914 | 0.911 |
| iWAR RA | 3.728 | 4.674 | −0.010 | 0.839 | 0.917 | 0.916 |

## S9.6. Added components and the compatible v0.23 overlap

**Table S21. AUGMENTED minus CORE under feature-specific development selection.** Same internal rows as Tables S9 and S15. Positive differences mean larger AUGMENTED error. New Ridge and HGB point contrasts lack stored matching intervals; unavailable does not imply equality.

| Target | Family | N | CORE RMSE | AUG RMSE | Difference | 95% interval |
|---|---|---|---|---|---|---|
| F | MLP | 1552 | 9.999 | 10.163 | +0.165 | [−0.056, 0.467] |
| F | Ridge | 1552 | 9.935 | 9.946 | +0.012 | Unavailable |
| F | HGB | 1552 | 10.119 | 10.138 | +0.019 | Unavailable |
| R | MLP | 1558 | 4.253 | 4.282 | +0.029 | [−0.088, 0.122] |
| R | Ridge | 1558 | 3.938 | 3.848 | −0.089 | Unavailable |
| R | HGB | 1558 | 4.323 | 4.639 | +0.315 | Unavailable |

A compatible v0.23 fWAR replay subset was established from actual saved predictions: all 504 new four-WAR pairs matched franchise key, target year, target games, credited label and fixed fWAR prediction exactly. On that subset the fWAR RMSE is 11.291, compared with 10.316 for Ridge AUGMENTED and 10.297 for neural AUGMENTED. The original 1,441-pair, 78-year fWAR RMSE of 11.076 is not a paired comparator for the full 1,552-row neural study. v23_shared_comparison.csv gives the supported direct subset comparison. These are publication-stage point estimates; their sparse historical membership does not support transferring full-cohort intervals.

## S9.7. Complete saved results, sensitivities and resource scope

publication_metrics.csv contains the internal, four-WAR and broader-bWAR tables for the full evaluation and modern 2010–2025 subset, each fixed/calibrated mapping separately. publication_contrasts.csv contains exact paired effects and interval eligibility. component_contrasts.csv contains family-specific component contrasts, including modern/era descriptions where supported. The full saved_neural_metrics.csv and

saved_neural_paired_effects.csv preserve every originally exported stratum and sensitivity: 1970–1989, 1990–2009 and 2010–2025 eras; equal-year weighting; actual-season credited-win units; target games at least 140 (both source and target at least 140 for F); bounded/clipped predictions; each neural seed; and leave-one-target-year-out influence summaries in the accompanying saved files. These machine-readable results form part of the supplement; no subset was dropped because its ranking was unfavorable. The sensitivity, weighting, unit, stratum and method columns in saved_neural_metrics.csv and saved_neural_paired_effects.csv preserve the available clipping, equal-year, actual-season, era, short-season and individual-seed results; saved_neural_leave_one_year_out.csv records the influence summaries. Field names and exact file inventory are documented with the released data dictionary. The original primary comparator stays fixed despite descriptive alternatives ranking better on some subsets.

The completed computational campaign recorded 2,975 completed invocations plus one failed/interrupted attempt (2,976 charged invocations). Invocations are fitting/diagnostic calls, not independent experiments. Recorded active measured intervals totaled 40.41 minutes; the conservative budget-charged total was 45.38 minutes, including documented allowances and recovery accounting. Sampled process-tree RSS reached 838,000,640 bytes (0.780 GiB), sampled every 0.25 seconds, rather than continuously measured peak memory. The machine was an AMD Ryzen 7 Pro 7735U with approximately 31.313 GiB RAM; one worker used two CPU threads. No rented compute was used. The retained original resource ledger defines inclusion and excludes this later publication reporting; new saved-only reporting time is recorded separately in evidence_receipt.json and the publication verification receipt.

### S9.8. Related work and contribution

The accompanying related_work.csv and RELATED_WORK.md inventory fourteen sources with target, unit, information timing, split, neural and non-neural comparators, publication status and access level. Hall and Seaman (2009), Farrell (2019; student project) and Lee et al. (2020)

establish earlier neural team-performance/win studies. Fort, McMichael and Rucker (2025) directly address team-level bWAR/fWAR and wins production; only publisher abstract and section snippets were accessible for closeout, so no claim about their uninspected full validation is made. openWAR (Baumer, Jensen and Matthews, 2015) and Grid WAR (Brill and Wyner, 2024) are established valuation precedents. The other game-outcome studies address different information sets or targets; completed-game recovery of a winner cannot be read as pregame forecast accuracy. The sequence-model study by Ahn et al. (2026) remains a preprint addressing another scale and task. No priority or state-of-the-art claim follows from this inventory. The contribution here is the transparent historical implementation, explicit exclusions and matched comparisons across targets, conversions and source coverage, together with bounded amended chronological learning evidence. The supplementary reference list includes the sources cited here; the supplied complete BibTeX and access table preserve the remaining inventory entries without placing unrelated accuracy percentages in the empirical leaderboard.

### S9.9. Reproduction and evidence boundaries

The v1.0.0 release archived at https://doi.org/10.5281/zenodo.23455100 supplies the actual conventional and neural scientific implementations, adapters, configurations, protocols and selected meaningful existing tests. Public aggregate reporting uses only released aggregate CSV/JSON files and figure sources. Source-level fitting requires authorized Retrosheet/provider observations, genuine retained membership/feature inputs and the original environment and freeze/ledger checks. Those inputs, exact row predictions, resampling indices and model objects are retained locally and excluded from public/review aggregates. No scientific fitting mode was executed in this closeout. Main and supplementary numerical displays are generated from publication_data.json; complete saved metrics remain full precision. The original conventional benchmark tables and native mathematical definitions were preserved. The code is also available at https://github.com/MRDOANE/iwar-historical-benchmark.

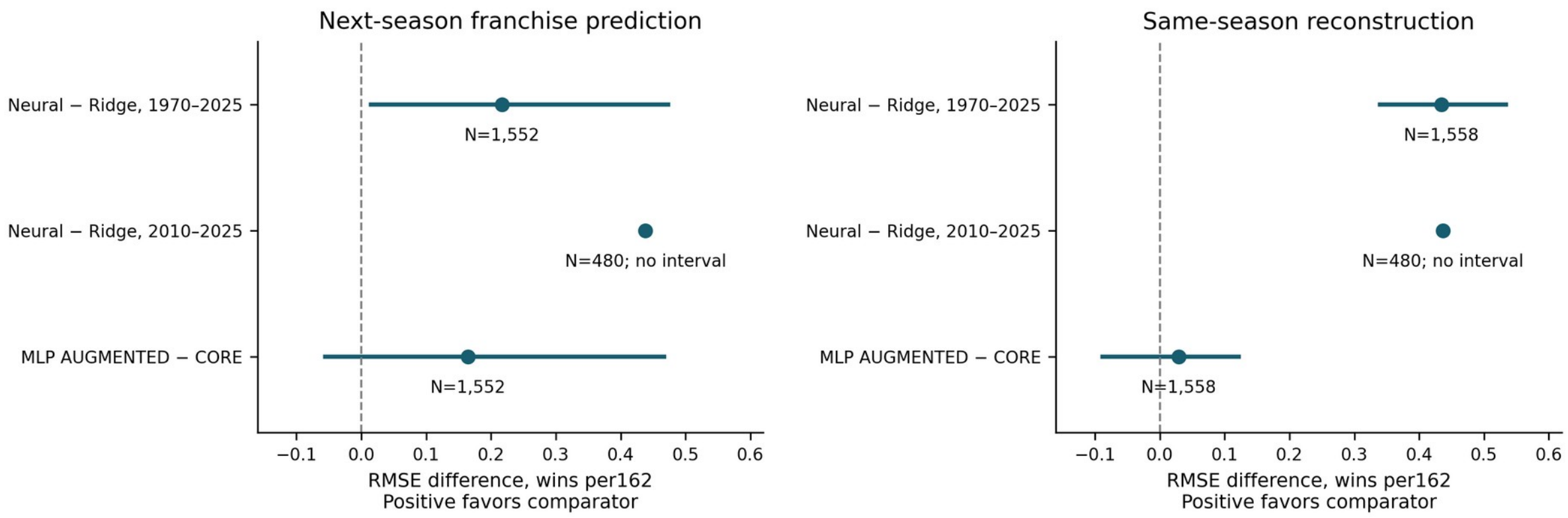

Figure S1. Saved neural-minus-Ridge and neural AUGMENTED-minus-CORE RMSE differences. Units are credited wins per 162; positive values mean larger neural or AUGMENTED error. Full cohorts contain 1,552 replay pairs and 1,558 reconstruction rows; modern cohorts contain 480 rows each. Lines are exact saved conditional pointwise 95% intervals with fitted predictions held fixed. Modern intervals are omitted under the ten-block rule. No new resampling was performed.

## Supplementary references